\documentclass[iop,apj,twocolappendix]{emulateapj}
\pdfoutput=1 %for arXiv submission
\usepackage{amsmath,amssymb,amstext}

\usepackage[breaklinks,colorlinks,citecolor=blue,linkcolor=blue]{hyperref} 

\usepackage[all]{hypcap} %Links go to figures; breaks on deluxetables (use \capstartfalse \capstarttrue to fix it)

\usepackage{aas_macros}
\usepackage{natbib}

\defcitealias{2016ApJ...819...29B}{BuZD2016}

\usepackage[table]{xcolor}
\shorttitle{The Fate of Supernova-Heated Gas in the LMC}
\shortauthors{Bustard et al.}

\begin{document}

\title{The Fate of Supernova-Heated Gas in Star-Forming Regions of the LMC: Lessons for Galaxy Formation?}

\author{
Chad Bustard\altaffilmark{1}, Stephen A. Pardy\altaffilmark{2}, Elena D'Onghia\altaffilmark{2,3}, Ellen G. Zweibel\altaffilmark{1,2}, J. S. Gallagher III\altaffilmark{2} 
}
\altaffiltext{1}{Physics Department, University of Wisconsin-Madison, 1150 University Avenue, Madison, WI 53706; bustard@wisc.edu}
\altaffiltext{2}{Department of Astronomy, University of Wisconsin-Madison, 2535 Sterling Hall, 475 N. Charter Street, Madison, WI 53706}
\altaffiltext{3}{Center for Computational Astrophysics, Flatiron Institute, 162 Fifth Avenue, New York, NY 10010, USA}

\begin{abstract}
Galactic winds and fountains driven by supernova-heated gas play an integral role in re-distributing gas in galaxies, depositing metals in the circumgalactic medium (CGM), and quenching star formation. The interplay between these outflows and ram pressure stripping due to the galaxy's motion through an ambient medium may enhance these effects by converting fountain flows into expelled gas. In this paper, we present controlled, 3D simulations of ram pressure stripping combined with thermally driven, local outflows from clustered supernovae in an isolated disk galaxy modeled on the Large Magellanic Cloud (LMC), a dwarf satellite of the Milky Way on its first infall. Observational evidence of local outflows emanating from supergiant shells in the LMC and a trailing filament of HI gas originating from these regions -  with no obvious Leading Arm counterpart - may represent a perfect example of this process. Our simulations present a proof-of-concept that ram pressure can convert fountain flows into expelled gas. We find that fountains launched near the peak star formation time of the LMC can comprise part of the LMC filament in the Trailing Stream, but with lower column densities than observed. Larger, more numerous outflows from the LMC may be possible and may contribute more mass, but higher inertia gas will lengthen the timescale for this gas to be swept away by ram pressure. Given the high resolution observations, increased knowledge of star formation histories, and growing evidence of multiphase, ionized outflows, the LMC is an ideal test-bed for future wind models.  

\end{abstract}


\section{Introduction}
\label{introduction}
There are a number of processes that remove gas from galaxies. These include environmental effects, such as ram pressure stripping (RPS) from the galaxy's motion through its background medium \citep{1972ApJ...176....1G} and tidal stripping resulting from gravitational interactions with neighboring objects, as well as internal feedback mechanisms such as supernova-driven outflows. The interplay between these mechanisms, which all play a role in the expulsion of gas from galaxies, is important to understand.

Supernova-driven outflows driven by a combination of thermal energy \citep{1985Natur.317...44C, 2009ApJ...704..137J}, cosmic ray energy \citep{1991A&A...245...79B, 2008ApJ...674..258E, 2014MNRAS.437.3312S, 2017ApJ...834..208R}, and radiation pressure on dust grains \citep{2005ApJ...618..569M, 2011ApJ...735...66M} are prevalent in galaxies spanning a large range of masses and represent a crucial process in galaxy formation and evolution \citep{2005ARA&A..43..769V, 2012MNRAS.421.3522H, 2014MNRAS.445..581H}. Galactic winds and fountains can pollute the CGM with metals, can inhibit future star formation by removing gas from central star-forming regions, and can transfer gas between galaxies. In low-mass dwarf galaxies, such as the LMC, these processes may be amplified due to shallow gravitational potentials that allow easier expulsion of gas as long as star formation and supernovae are efficient enough to provide adequate wind launching. Indeed, simulations of isolated disk galaxies show evidence for preferentially higher mass-loading in winds from low-mass galaxies \citep{2016ApJ...824...57C}, and cosmological simulations require efficient stellar feedback at low masses to account for the observed stellar mass - halo mass relation \citep{2010ApJ...710..903M, 2014MNRAS.445..581H}. Much of this outflow gas seems to recycle back onto the galaxy \citep{2015MNRAS.454.2691M, 2016ApJ...824...57C, 2017MNRAS.470.4698A, 2017MNRAS.468.4170M}, possibly residing in the CGM of the host galaxy for hundreds of Myrs to a few Gyrs before falling back to the disk, typically with a larger angular momentum after being ``spun-up" in the halo \citep{2012MNRAS.419..771B}.

In less energetic galaxies with lower star formation rates, such as some dwarf galaxies of the Local Group, intense radiative cooling can inhibit wind launching even for these systems with shallow gravitational potential wells \citep{Emerick2016GASSATELLITES}. Within about 250 kpc of the Milky Way, however, all dwarf satellite galaxies (with the exception of the Magellanic Clouds) are undetected in HI \citep{2009ApJ...696..385G} and devoid of observable star formation \citep{2012AJ....144....4M}. This correlation with distance from the Milky Way suggests that ram pressure stripping, which becomes stronger with increasing halo density and relative velocity close to the Galaxy, may be significant.  

In many systems it is likely that neither of these individual mechanisms dominate, and a combination of stripping processes is needed to account for the observed gas loss. One example is the interplay between ram pressure and galactic ``fountains'' \citep{1976ApJ...205..762S, 1995ApJ...440..634R}, whereby supernova-heated gas is launched above the disk but ultimately falls back onto the galaxy. In this paper, we study this interplay between RPS and galactic fountains as a proof-of-concept that ram pressure can convert a galactic fountain flow into expelled gas, thereby providing a gas-loss mechanism more efficient than ram pressure stripping or fountain flows acting individually. 

Our controlled simulations complement previous works on this subject \citep{1999MNRAS.309..161M, 2004MNRAS.352..363M}, which model the effects of galactic winds in tandem with ram pressure stripping. The outcome depends on the surrounding environment and the structure and gravitational potential of the system. While RPS can deposit high pressure gas in the galaxy's halo and subsequently limit superbubble breakout from the disk, it can also sweep away outflows that do propagate far enough above the disk. Crevices that form in the disk from such blowouts can make RPS more efficient due to the more complex density structure on which that ram pressure acts, and the effective thickening of the disk from supernova-driven outflows may cause the galaxy to respond more impulsively to tidal forces \citep{2017ApJ...836L..13K}. Such effects may be secondary for galaxies already being very rapidly disintegrated by strong RPS; however, for galaxies undergoing weaker RPS, the combination of RPS and outflows may help account for the observed gas loss. High mass dwarfs of the Local Group, for example, which are not as easily stripped due to their higher escape velocities, may have sufficient star formation to launch outflows and utilize these effects. In fact, ram pressure may actually ignite such a mechanism - modest amounts of ram pressure can compress the interstellar medium and trigger star formation. This provides two gas loss mechanisms: 1) some gas is simply sunk into stars and 2) these newly formed stars may lead to supernovae and wind driving, at which point ram pressure can sweep away the loosely bound gas that would otherwise be trapped within the disk. This highlights the dual role of ram pressure in both promoting and impeding star formation, depending on the parameter regime.

We specifically apply these ideas to the Large Magellanic Cloud (LMC), a dwarf satellite of the Milky Way and one of the best-studied objects in extragalactic astronomy. As described below, certain morphologies of the Magellanic system, specifically the LMC filament, may be examples of this exact gas-loss mechanism. 

There is increasing evidence that the LMC possesses local, supernova-driven winds. High-resolution HI observations of the LMC show that the almost face-on disk is punctured with prominent holes \citep{2003MNRAS.339...87S}. Coincident with these holes are %extensively studied, 
x-ray emitting supergiant shells (SGSs) \citep{1980MNRAS.190..403M, Kim1999HICloud}, suggesting that these once star forming regions were carved out by supernova heating and possibly local galactic outflows. Most recently, \citet{2016ApJ...817...91B} present considerable evidence for $\approx 100$ km/s photoionized outflows from the LMC, which may originate from such regions. Combined with an expanding knowledge of star formation histories \citep{2008PASA...25..116H, Harris2009THECLOUD, Piatti2017StarCloud}, the well-studied physical structure of the LMC and evidence for outflows make the LMC an interesting test-bed for wind launching.   

Given that many of the HI holes are located along an arc on the leading edge of the LMC disk, one scenario that could lead to intense regions of star formation is compression of the leading edge of the LMC as it moves through the hot halo of the Milky Way \citep{2009MNRAS.399.2004M}. In this case, regions that rotate into the leading edge would undergo increased star formation and subsequent supernovae. Gas at these sites would be heated, consequently showing a hole in HI observations but filled with x-ray emission \citep{1994A&A...283L..21B, 1999ApJ...518..298P, 2000ApJ...545..827P, 2001ApJS..136...99P}. The hole will eventually collapse as it rotates around the galaxy; however, while the pressure difference between the hole and its surroundings is sustained by supernovae, a local, thermally driven outflow may be driven above the disk before the hole collapses. This leads to an interesting question: \emph{If an outflow does occur in the LMC, where does the expelled gas go? }

Recent results from the FIRE cosmological simulations suggest that winds launched from dwarf galaxies may represent a significant mode of intergalactic gas transfer onto larger, central host galaxies \citep{2017MNRAS.470.4698A}. This may fuel the host galaxy along with other accretion and merger events. \citet{Indu2015AstronomyOutflow} study the observed HI kinematics of the LMC and suggest that some well-known features (e.g. Arm E on the leading edge) may actually be outflows. Outflows may also contribute to the Magellanic Stream, which is believed to be caused by tidal stripping between the Small Magellanic Cloud (SMC) and the LMC \citep{2012MNRAS.421.2109B}, although the total mass of the Stream is underestimated (see \cite{2016ARA&A..54..363D} for a recent review). This gaseous trail behind the Clouds could supply the Milky Way with a large amount of gas if it can fall into the Galactic disk, making it an intriguing, nearby case-study for galaxy evolution and the so-called ``baryon cycle'' that regulates star formation and galaxy growth through gas expulsion, accretion, and galaxy merger events. 

There are two filaments that make up the Stream; one originates in the SMC, while another points back to a portion of the LMC known as the Southeast HI Overdensity (SEHO) \citep{Nidever2008TheArm}. This quadrant includes the intensely star-forming 30 Doradus region, which will almost certainly undergo a significant period of supernovae and possible mass ejection in the near future. \citet{Nidever2008TheArm} speculate that the LMC's numerous HI holes were once very similar to the 30 Doradus region and blew out a significant amount of material. As this material was swept up by the surrounding velocity field due to the LMC's motion through the Milky Way halo, it may have stretched across the LMC disk and taken on the appearance of the LMC filament when viewed in almost face-on projection. This scenario is especially interesting because the observed LMC filament is comprised almost entirely of gas and devoid of stars \citep{1982MNRAS.201..473R, 1983A&A...124..216B, 1998ASPC..136...22G}. If this filament were formed from tidal stripping alone, one would expect more stars to be present in the filament. Sinusoidal variations in the filament, both spatially and in velocity, may be an imprint of the rotation of the galaxy and/or temporal separation of the outbursts. 

Two prominent shells are SGS4 and SGS12. Comparing column densities inside and around the holes, the HI column density absent from the holes is on the order of $10^{21} \rm cm^{-2}$. This implies an HI mass loss of $5.65 \times 10^{5} M_{\odot}$ for SGS12 and $1.06 \times 10^{5} M_{\odot}$ for SGS4. For comparison, the total HI mass of the Magellanic Stream is $> 10^{8} M_{\odot}$ \citep{2005A&A...432...45B} with more than three times that mass in ionized gas \citep{2014ApJ...787..147F}. This suggests that, if the Stream was formed entirely by these outflows, on the order of a few hundred to a thousand of these shells would have had to blow out \citep{Nidever2008TheArm}. It is more likely that a combination of outflows and other processes such as tidal stripping account for the Stream, though these outflows should not be overlooked, especially regarding formation of the LMC filament.

Recent numerical simulations show that if the Clouds were more gas rich and more extended in their HI disk, then they would likely create a bifurcation in the Trailing Arm with one filament structure belonging to the LMC and one to the SMC. However, the same filaments would appear in the Leading Arm \citep{2018arXiv180201600P}. Recent observational work on the metallicity of the Leading Arm indicates that the gas metallicity is lower than the LMC and the SMC today and more consistent with the metallicity of the SMC a few Gyrs ago \citep{2018arXiv180106446F}. 

In this paper, we model individual outflows from the LMC disk and their contribution to the Stream via RPS. We present a suite of 3D simulations using the FLASH hydrodynamics code \citep{2000ApJS..131..273F}, including the disk geometry of the LMC (described in Section \ref{galaxymodel}), radiative cooling, RPS (described in Section \ref{rammodel}), and time-dependent wind launching via thermal energy and mass injection from supernovae (described in Section \ref{windlaunching}). We choose in this work to only consider thermal energy injection as the wind driving mechanism. In reality, supernova energy deposition is split between 
%some fraction of 
thermal, kinetic, and cosmic ray energies. Purely thermal winds, because they lose much of their pressure support to radiative losses, represent the least optimistic scenario for wind launching. For now, we leave injection via all three modes of energy injection to future work. In Section \ref{windandram_section}, we present proof-of-concept simulations of ram pressure expelling fountain gas from the mock LMC galaxy. Individual outflows launched near the peak star formation time can contribute blobs of gas tens of kpcs behind the galaxy. Sequentially launching multiple winds can lead to a more filamentary structure but of slightly less spatial extent. In either case, more numerous or larger outbursts are required to account for the observed column densities in the LMC filament. In Section \ref{Conclusions}, we expand on these conclusions and speculate on the broader impacts for galaxy formation and feedback in dwarf galaxies.

\section{3D Galaxy Model}
\label{galaxymodel}
In this section, we describe our setup of a mock LMC disk galaxy using the adaptive mesh refinement (AMR) code FLASH 4.2 \citep{2000ApJS..131..273F}. We use the directionally unsplit staggered mesh hydrodynamics solver \citep{2013JCoPh.243..269L}. For this work, no magnetic field is included, though we note that magnetic fields may naturally form ram pressure stripped material into filaments \citep{2014ApJ...784...75R}, which could help create the LMC filament. 

Our galaxy is placed in a cartesian grid with $\hat{x}$ and $\hat{y}$ in the galaxy midplane and $\hat{z}$ pointing out of the plane. In isolated wind simulations without ram pressure, our base grid is 30 kpc x 30 kpc x 20 kpc with 256 cells in each direction. This represents a resolution of 78 pc in the vertical direction. Simulations with this level of resolution are referred to as ``lowRes.'' In the wind injection region, AMR is utilized with 2 additional levels of refinement. This gives us a maximum resolution of 19.5 pc, which we refer to as ``medRes.'' With wind injection occurring on scales of order 100 pc, energy and mass are effectively added into a few hundred surrounding cells at each timestep. A small resolution study is presented in Section \ref{resolution} including a wind run with 3 levels of refinement (``highRes''). Section \ref{discussion} expounds on the numerical limitations as well as the physical limitations of our model. 

In simulations with ram pressure included, our grid sits in the frame of the galaxy, which moves in the $-\hat{x}$ direction, meaning that ram pressure affects the disk from left to right. We extend the x-direction to 60 kpc long and the y-direction to 40 kpc long in order to follow the stripped and outflowing material as it moves behind the galaxy. A specialized refinement routine is used in this case to better resolve the cells within 25 kpc of the galaxy center, while the refinement drops one level for cells further away. For computational feasibility, no extra refinement is used near the outflow launch site, and the maximum resolution is 78 pc near the galaxy. One additional AMR simulation with a maximum vertical resolution of 19.5 pc, corresponding to a ``medRes'' simulation, is compared in Section \ref{RPS_resTest} to the ``lowRes'' run. More mass is expelled from the disk in this case, but the overall conclusions on the ram pressure - fountain interaction and the formation of the LMC filament aren't significantly changed. Future high resolution simulations with spatially resolved momentum and energy injection may shed further light on this effect. 

\subsection{Galaxy Setup}
\label{initialsetup}
In setting up an equilibrium LMC disk, we take inspiration from \cite{Tonnesen2009GASMEDIUM} (also see \citealt{Roediger2006RamAngle}, \citealt{2009MNRAS.399.2004M}, \citealt{Salem2015RAMMEDIUM}), which considered the effects of RPS on disk galaxies. Specifically, there are three elements of our setup that we deem to be most important, especially for simulations involving outflows. 

1) The gravitational potential of the system is important for counteracting both outflows and stripped material. We model the stellar potential of the disk as a static Plummer-Kuzmin disk \citep{1975PASJ...27..533M} potential:
\begin{equation}
\Phi(R,z) = GM_{\star} \left[R^{2} + \left(a_{\star} + \sqrt{b_{\star}^{2} + z^{2}} \right)^{2} \right]^{-1/2}
\end{equation}
We use a total stellar mass of $3 \times 10^{9} M_{\odot}$, a radial scale length of $a_{\rm gas} = 1.7$ kpc and a vertical scale height of $b_{\rm gas} = 0.34$ kpc (see Table~\ref{propsTable} for all initial galaxy parameters.) These parameter choices are motivated by \citet{Salem2015RAMMEDIUM}, which found a good fit to the observed shape of the LMC disk. We note that, for simplicity, we do not explicitly include star particles in our simulation. They only effectively act on the gas through the imposed, external potential. We also neglect the potential from the dark matter halo, which is likely negligible in our region of interest within a few kpc from the center of the disk. From our steady-state results, we also find that halo mass has a negligible effect on the asymptotic wind velocity (See Figure 6 of \citealt{2017ApJ...835...72B}.) In addition, it is possible that the dark matter halo has been lost due to tidal stripping with the SMC \citep{2004ApJ...609..482K}. We also leave the effects of self-gravity, which should be small given the almost order of magnitude difference between the stellar and gas masses of the disk, to future work. 

2) The vertical gas profile of the disk should be realistic. This is important because the thermal pressure gradient developed by supernova energy injection needs to act on the correct vertical gas column. If the disk is too thin, too little gas will be ejected above the disk, whereas if the disk is too thick, the outflow can be either over-dense above the disk or too heavy to be lifted upwards by a thermal pressure gradient at all. We set up our gas disk as a standard hyperbolic secant profile, which converges to an exponential disk at large heights and radii. Specifically, following \citet{Tonnesen2009GASMEDIUM},

\begin{equation}
\rho (R,z) = \rm \frac{M_{gas}}{2\pi a_{gas}^{2} b_{gas}} 0.5^{2} \rm sech\Big(\frac{R}{a_{gas}}\Big) \rm sech\Big(\frac{|z|}{b_{gas}}\Big) 
\end{equation}
To avoid complications at the boundaries, we truncate the disk by multiplying the density by a factor $0.5(1+cos(\rm \pi (R - 10 kpc)/13 kpc))$ for radii between 10 and 13 kpc. We also truncate the density at a floor of $10^{-29} \rm g/cm^{3}$, which represents the density of the surrounding intergalactic medium \citep{2017arXiv171002116L}.

\begin{table}[htb]
\centering
\label{propsTable}
\begin{tabular}{ c | c }
Variable & Value \\
\hline
$M_{\star}$  & $3 \times 10^{9} M_{\odot}$  \\
$a_{\star}$  & 1.7 kpc \\
$b_{\star}$  & 0.34 kpc  \\
$M_{\rm gas}$ & $5 \times 10^{8} M_{\odot}$ \\
$a_{\rm gas}$ & 1.7 kpc \\
$b_{\rm gas}$ & 0.34 kpc  \\
\end{tabular}
\caption{Table of galaxy stellar and gas parameters. $M_{\star}$ and $M_{\rm gas}$ are the stellar mass and gas mass, respectively. $a_{\star}$ and $a_{\rm gas}$ are the stellar and gas scale lengths, which are assumed here to be equal. $b_{\star}$ and $b_{\rm gas}$ are the stellar and gas scale heights, which are assumed to be one-fifth of the scale lengths.}
\end{table}

The pressure profile is chosen so that the pressure gradient balances the imposed gravitational force in the z-direction. Our gas is then given a rotational velocity to counteract the remaining difference between the radial gravitational force and the radial pressure gradient. We note that this configuration is stable in the absence of ram pressure and wind launching for at least a few hundred Myrs. At later times, small numerical perturbations do throw the disk slightly out of equilibrium, resulting in instabilities at the disk-IGM interfaces. This is likely due to our inability to properly resolve the pressure scale height, even at fairly high resolution. 

3) Radiative cooling is important for the wind launching and propagation, and it may also affect RPS. Therefore, we elaborate on our implementation of radiative cooling in the following section.

\subsection{Optically Thin Cooling}
\label{cooling}

\begin{figure}
\label{CloudyCurve}
\centering
\includegraphics[width = 0.49\textwidth]{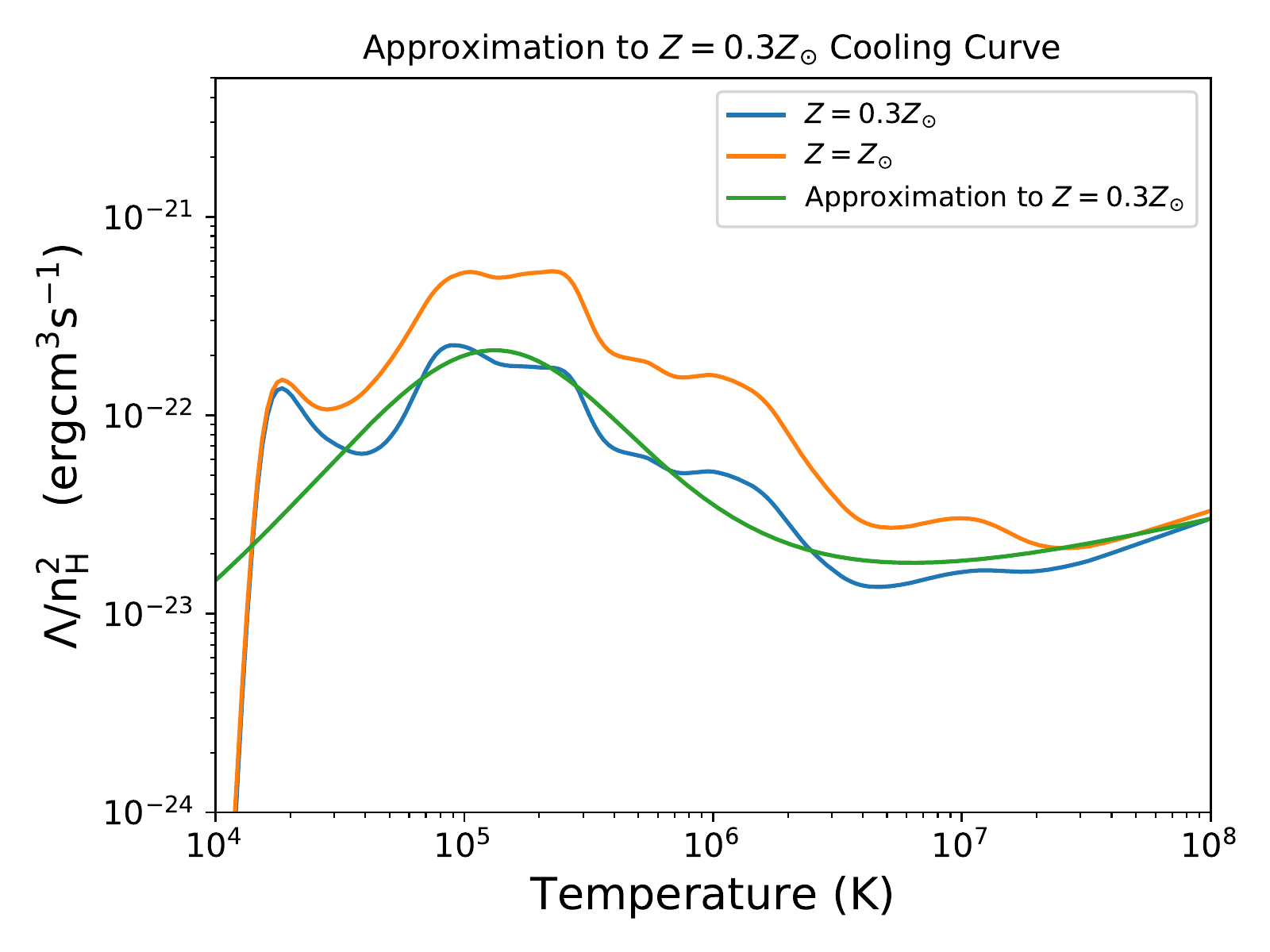}
\caption{Cooling curves made with CLOUDY as a function of temperature for abundances of $0.3 Z_{\odot}$ and $1.0 Z_{\odot}$. Also shown is the approximation we use to the $0.3 Z_{\odot}$ curve.}
\end{figure}

Radiative cooling can efficiently keep thermally driven winds from launching and also lead to sharp temperature decreases in heavily mass-loaded flows \citep{2003ApJ...590..791S, 2016ApJ...819...29B, 2016MNRAS.455.1830T}. Several studies, using the well determined ISM argon, oxygen, and sulfur abundances from HII regions, place the chemical abundance of the LMC as a whole (and 30 Doradus) at $Z \approx 0.3 Z_{\odot}$ \citep{2008ApJ...680..398L, Pellegrini2011STRUCTUREABUNDANCES}. In 30 Doradus in particular, \citet{2006AJ....131.2140T} used the spectral fitting package XSPEC to measure $0.3 Z_{\odot}$ for all elements except for O (abundance of 1.2 $Z_{\odot}$), Ne (abundance of $1.1 Z_{\odot}$), and Mg (abundance of $0.7 Z_{\odot}$).

These overabundances, particularly in oxygen, may significantly increase cooling compared to a case where all elements have $0.3 Z_{\odot}$. This inferred enrichment makes sense because the hot phase probed by \citet{2006AJ....131.2140T} consists of a mixture of ejecta from supernovae and the shocked interstellar medium (ISM); therefore, it can have a range of chemical abundances and is likely to be enriched in the alpha-elements by core-collapse supernovae. As a wind from such a region sweeps up material from the surrounding ISM, though, the enriched gas may mix with the lower abundance surroundings. For example, a highly mass loaded wind will be imprinted with the chemical composition of the surrounding ISM (e.g. \citealt{2010ApJ...710..948P}), while a hot phase produced by shocks between multiple SNe is likely to be enriched.

We choose for the rest of this paper to assume full mixing and create an approximation to just the $Z = 0.3 Z_{\odot}$ cooling curve generated using CLOUDY assuming collisional ionization equilibrium \citep{Wiersma2009ThePlasmas}. Figure \ref{CloudyCurve} shows our approximation, which is used in our FLASH models as well, and a comparison of $Z = 0.3 Z_{\odot}$ and $Z = Z_{\odot}$ cooling curves. Photoionization heating is not included in this work, although we note that this additional heating from the metagalactic UV background may offset some radiative cooling and crucially affect the outflow's ionization state and, hence, its observable properties, especially in the LMC halo. This is especially likely to be true at temperatures of 10$^4$K and below; in particular, we cannot follow the transition from H II to H I needed for quantitative comparison to observations of the filament. However, the temperature boost due to photoionization heating is expected to be far from adequate to drive gas out of the galaxy.

\begin{figure}[]
\label{tempFloor}
\centering
\includegraphics[width = 0.49\textwidth]{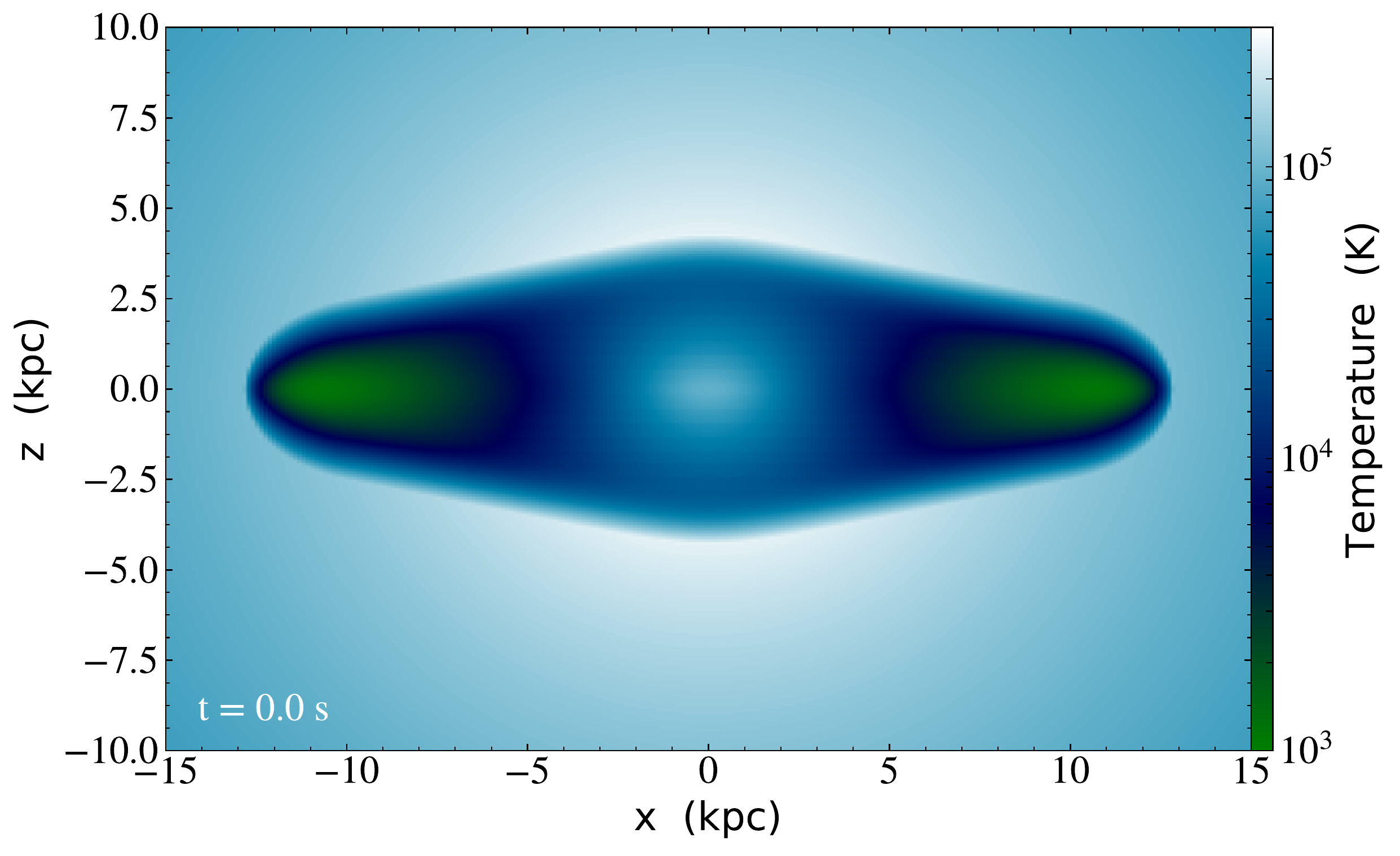}
\includegraphics[width = 0.49\textwidth]{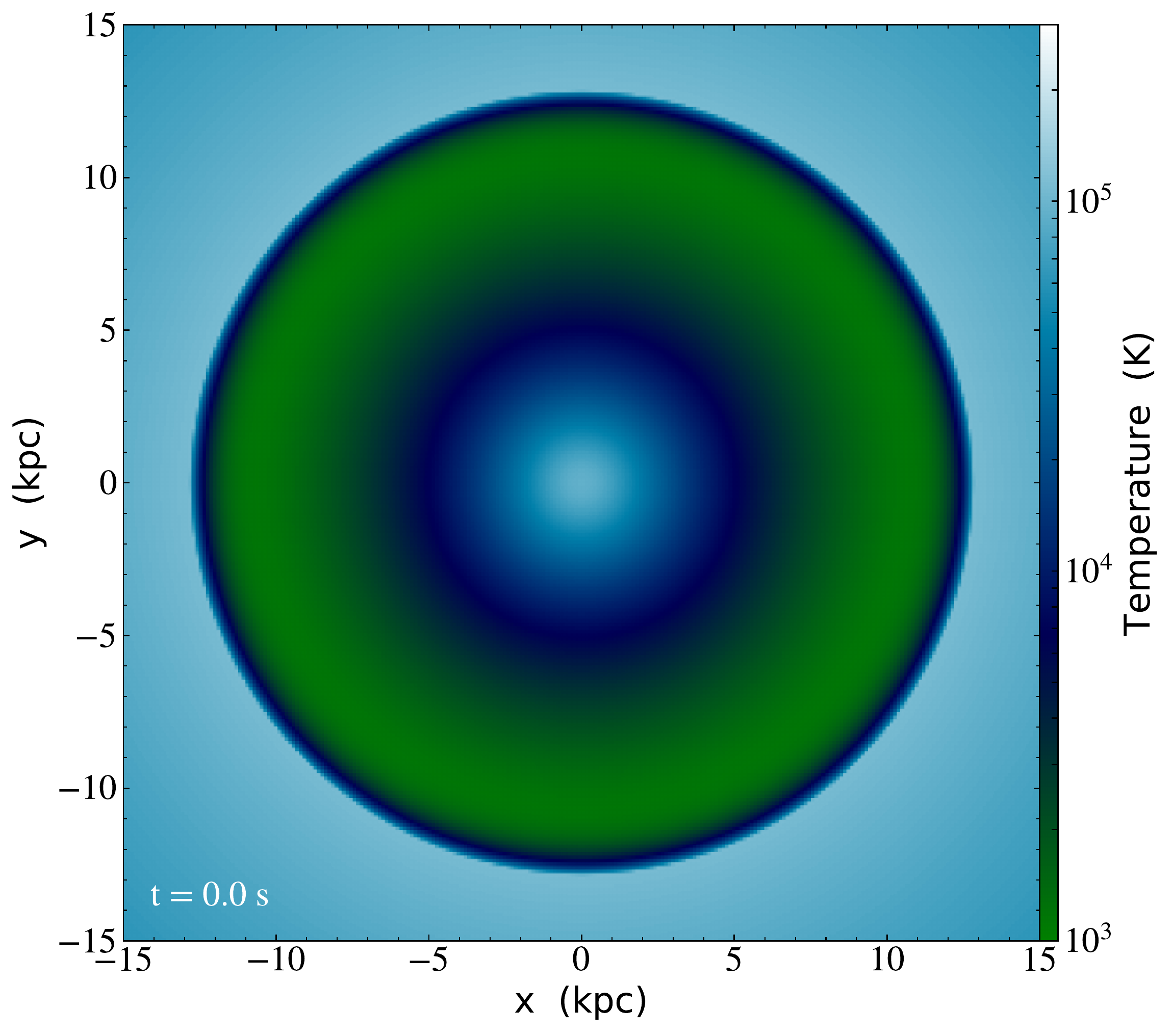}
\caption{Vertical (top) and horizontal (bottom) slices of initial temperature in our simulations. This initial temperature is used as a temperature floor to determine whether radiative cooling operates fully or not. This floor has no bearing on adiabatic cooling, though. Temperature, which we view as the combination of thermal and turbulent specific energy, ranges from $10^{3}$ K to a few times $10^{5}$ K. In the inner regions where outflows are presumed to be launched, the temperature is between $10^{4}$ and a few times $10^{5}$ K.}
\end{figure}

Our implementation of radiative cooling is necessarily delicate in that letting our mock galaxy cool fully would create pressure gradients directed towards the disk plane. This would throw our galaxy out of equilibrium and change the gas profile, which is important for the outflow launching and gas stripping. How turbulence and thermal pressure support a galaxy's scale height in tandem with cosmic ray and magnetic pressures is still an open problem, but the turbulent (random) velocity dispersion may be a factor of a few larger than the thermal velocity dispersion (see e.g. \citealt{2016ApJ...832..118B}). Solely thermal pressure at temperatures of $10^{4}$ K would provide inadequate disk support. Future work will include a disk with support from explicit turbulence, cosmic rays, and magnetic fields, as opposed to purely thermal support. For now, we view the ``temperature'' in our work as a combination of thermal and turbulent specific energy, which allows the model to have a thick disk at a temperature, which if purely kinetic, would imply strong radiative cooling. Shown in Figure \ref{tempFloor}, these gas temperatures range from $\approx 10^{3}$ K near the edges of the disk to $10^{5}$ K near the galaxy center. Outside the disk, the temperature reaches a few times $10^{5}$ K. For this reason, it is difficult to compare our simulations to maps of just HI where the gas is at temperatures of order $10^{4}$ K. 

To keep our galaxy in equilibrium while maintaining the effects of cooling on the wind, we employ a cooling temperature floor for each grid cell, and we set that floor to be the initial temperature of that cell at the start of our simulation. Unlike a typically-used temperature floor, where both adiabatic cooling and radiative cooling are opposed by an artificial injection of internal energy, in our setup, a temperature floor is used \emph{only if the temperature decrease is due to cooling}. Wind and stripped material are still allowed to adiabatically cool to arbitrarily low temperatures, but radiative cooling can only cool gas to the floor value. Let's define the initial cell temperature as $T_{\rm floor}$, the current cell temperature as T, and the updated cell temperature resulting only from energy addition and radiative losses as $T_{\rm new}$. Cooling operates fully on the gas, except in the following cases:

\begin{align*}
\rm if \quad T > T_{\rm floor} \quad \rm and \quad T_{\rm new} < T_{\rm floor}, \\
T_{\rm new} = \rm T_{\rm floor} \\
\\
\rm if \quad T < T_{\rm floor} \quad and \quad T_{\rm new} < T_{\rm floor}, \\
T_{\rm new} = T 
\end{align*}

In wind launching regions, this implies a minimum energy injection rate required to increase T beyond its floor value. Otherwise, if cooling exceeds heating to the extent that the temperature would fall below the floor value, the temperature is set equal to that floor. For modest energy injection rates, the temperature will never exceed this floor threshold, and the galaxy will stay in its initial state. We refer to this as an ``inadequate heating'' case. On the other hand, if a hot wind is launched and lifted out of the plane, cooling will turn on because the temperature floor is lower moving away from the disk plane. Therefore, this method captures the effects of radiative cooling on both outflow launching and propagation. 

\section{Ram Pressure Stripping}
\label{rammodel}
\begin{figure*}[]
\label{SalemRam_hotHalo}
\centering
\includegraphics[width = 0.4\textwidth]{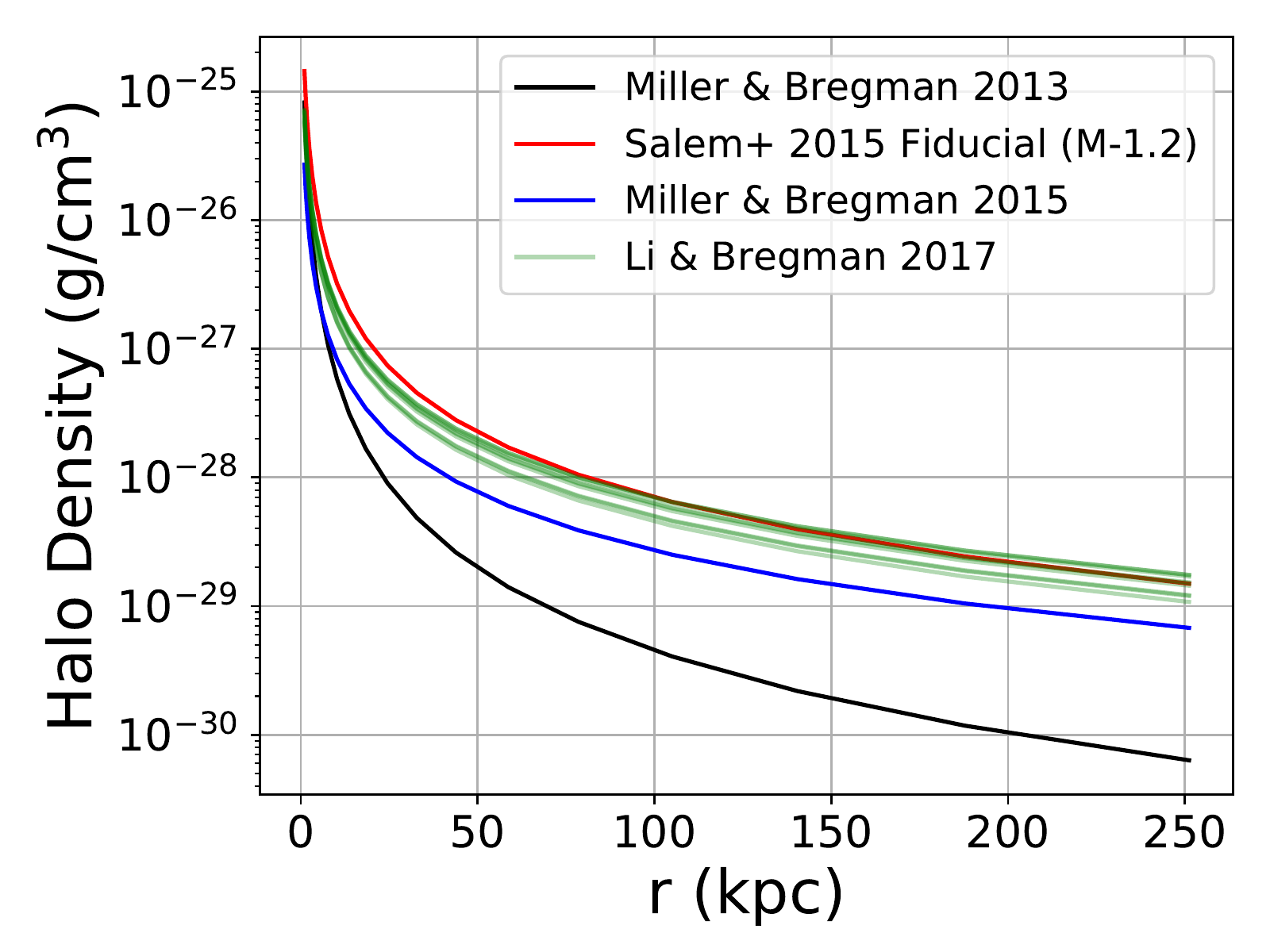}
\includegraphics[width = 0.4\textwidth]{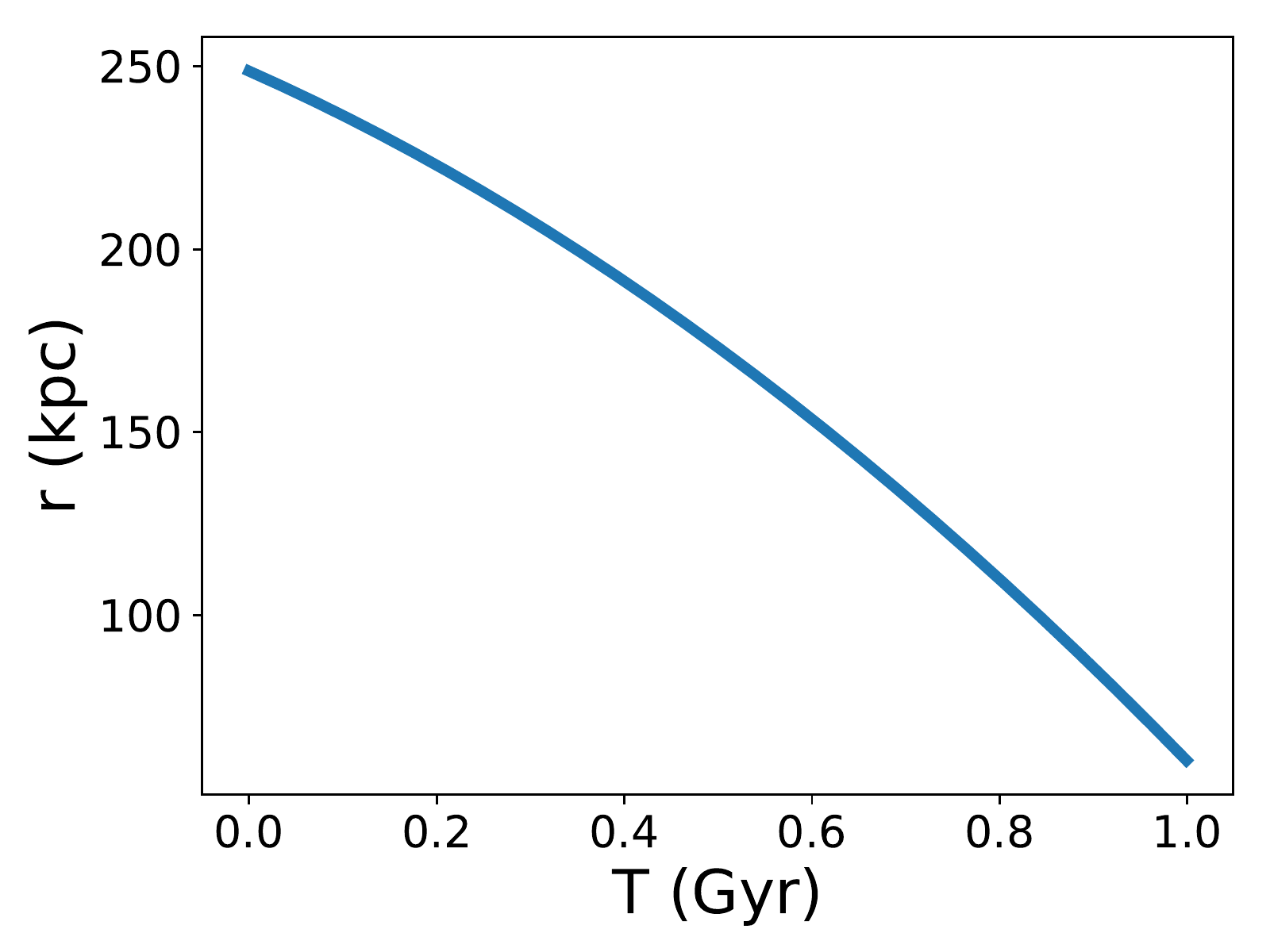}
\includegraphics[width = 0.4\textwidth]{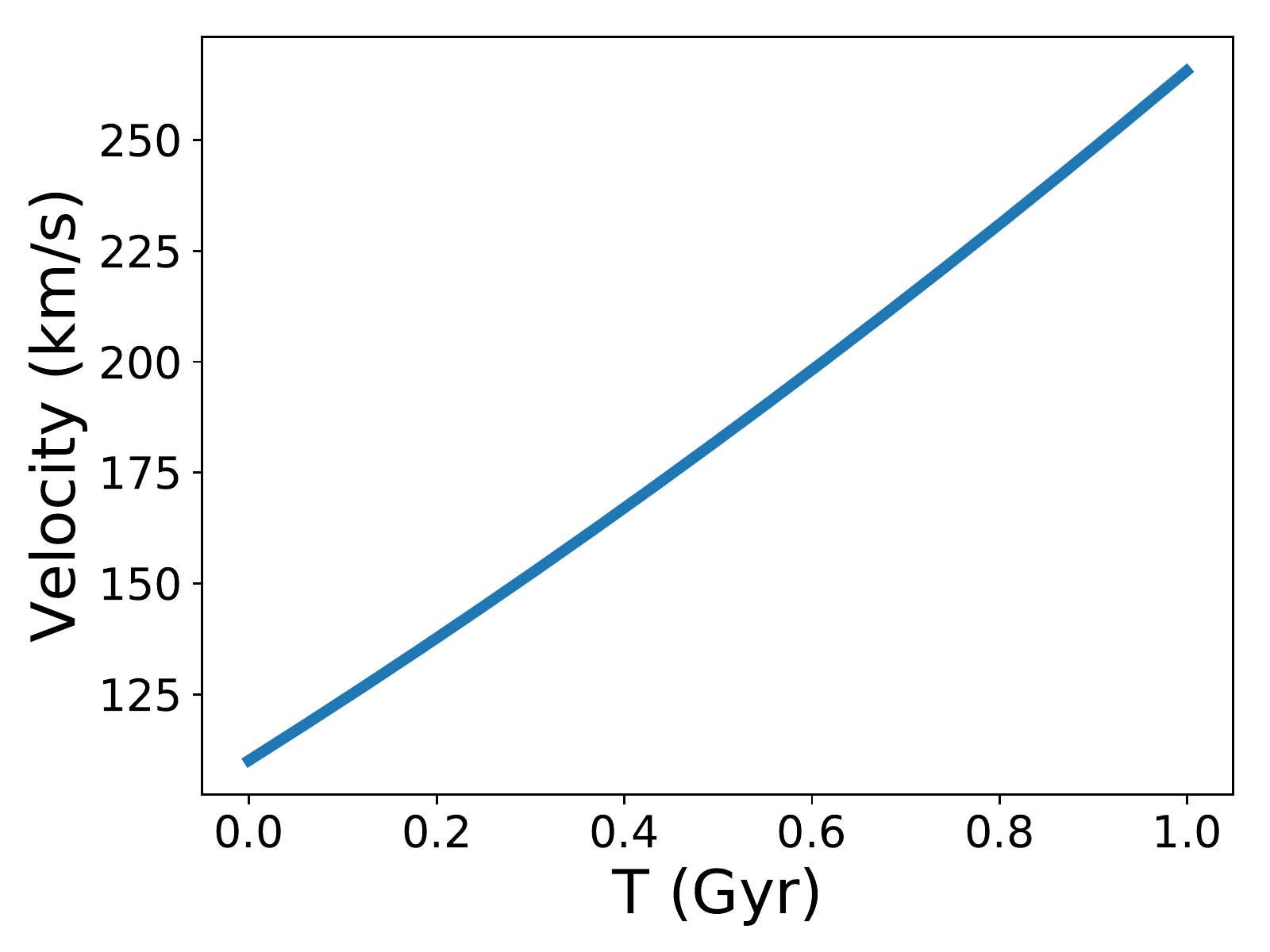}
\includegraphics[width = 0.4\textwidth]{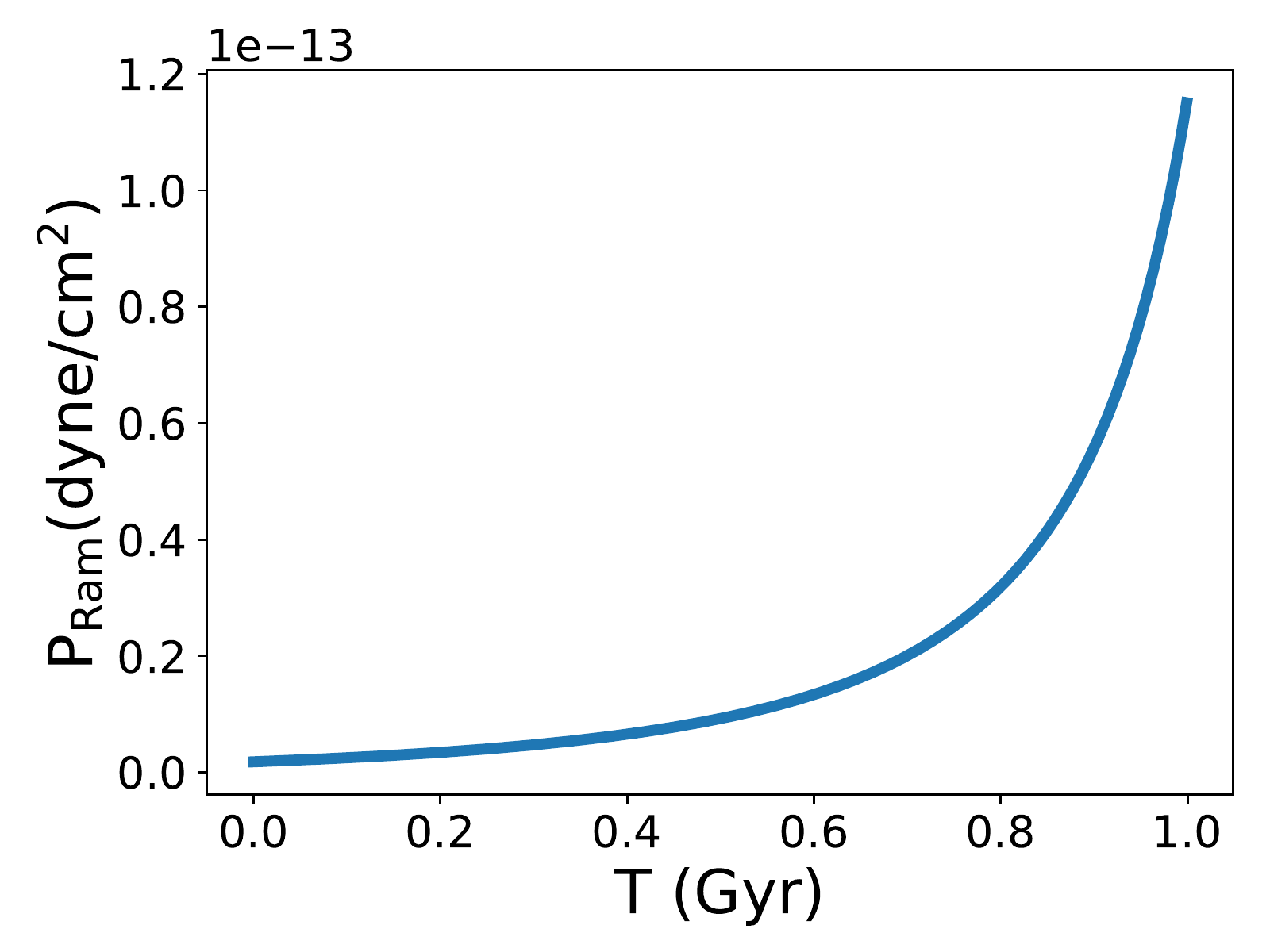}
\includegraphics[width = 0.9\textwidth]{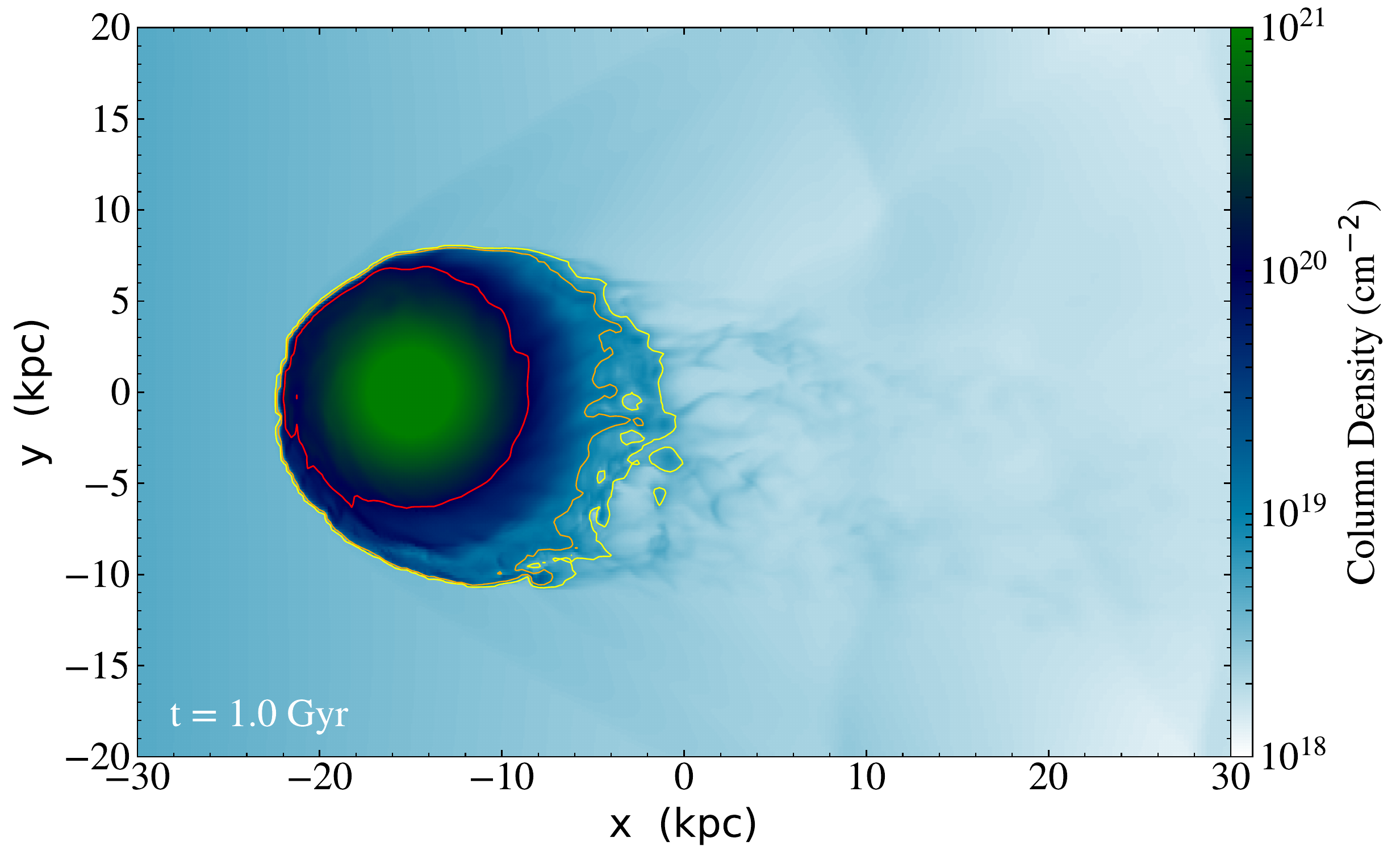}
\caption{Top: Best-fit $\beta$-profiles of the Milky Way halo density vs radius for three recent observational probes and the model of \cite{Salem2015RAMMEDIUM}. Also shown are the orbital radius and x-velocity of the LMC assumed in both this work and in \cite{Salem2015RAMMEDIUM}, as well as the total ram pressure as a function of time. Bottom: Present-day result of RPS after 1 Gyr using this ram pressure profile. Contours show projected face-on densities of $10^{20}$ (red), $10^{19}$ (orange), and $5 \times 10^{18} \rm cm^{-2}$ (yellow). As in \cite{Salem2015RAMMEDIUM}, the stripping is fairly weak and clearly unable to account for the entirety of the Stream, which shows much higher column densities tens of kpcs from the galaxy.}
\end{figure*}

For our ram pressure inflow, we utilize a wind tunnel boundary condition, which is standard practice for RPS simulations (e.g. \citealt{2005A&A...433..875R}, \citealt{Roediger2006RamAngle}, \citealt{2006MNRAS.369.1021M}, \citealt{2013MNRAS.433.2749G}). We slowly turn on this wind tunnel with a time profile of $1-e^{-t/\tau}$ and a characteristic timescale of $\tau = 50$ Myrs. The inflowing gas has a temperature of $10^{6}$ K, and the flow is only in the direction parallel to the plane of the disk (the x-direction). At present time, this is a fair assumption because the motion of the LMC is only mildly inclined relative to the disk plane. \citet{Roediger2006RamAngle} also showed that ram pressure inclination only significantly affects stripping when the inclination angle is greater than $\approx 60$ degrees relative to the disk midplane, which is not the case for the LMC. 

Because the ram pressure scales with the velocity squared, it is important to use a reasonable velocity profile. To calculate an orbit and expected velocity for the LMC we ran simplified models of the LMC and Milky Way interaction based on the proper motion measurements of \citet{2006ApJ...638..772K} and the initial positions and velocities of \citet{2012MNRAS.421.2109B}. At each timestep of the simulation we measured the x-component of the relative velocity of the LMC and the distance to the center of the Milky Way.

For the density of the Milky Way halo, we assume a $\beta$-profile \citep{1998ApJ...497..555M}, specifically,
\begin{equation}
\label{Makino}
n(r) = n_{0} \left[1+\left(\frac{r}{r_{c}}\right)^{2} \right]^{-3 \beta / 2} \approx n_{0} \left(\frac{r}{r_{c}} \right)^{3 \beta}
\end{equation}
and we assume the above approximation since our orbital region of interest is at large radii. We use the best fit parameters of $n_{0} = 0.46$ $\rm cm^{-3}$, $r_{c} = 0.35$ kpc, $\beta = 0.559$ from \citet{Salem2015RAMMEDIUM}, which used simulations of the LMC truncation radius as a measure of the hot halo density. A number of similar attempts have been made to probe the halo density through analytic RPS arguments \citep{2009ApJ...696..385G} and hydrodynamic simulations of RPS of other Milky Way dwarf galaxies \citep{2013MNRAS.433.2749G}. 

Although measuring the density profile is very challenging, observational determinations have consistently improved since the pioneering work of \citet{2013ApJ...770..118M} through x-ray O VII and O VIII emission and absorption lines, as well as UV OVI absorption lines \citep{2017ApJ...835...52F}. \citet{Salem2015RAMMEDIUM} find that their stripped LMC is best fit by a profile with a lower $\beta$ value, i.e. one in which the density is greater than that of \citet{2013ApJ...770..118M} at large distances, where the LMC has predominantly orbited throughout the past Gyr. Similarly, \citet{2015ApJ...800...14M} and \citet{2017arXiv171002116L} find ever-increasing densities at large radius. In the best-fit parameters of \citet{2017arXiv171002116L}, the density at large distances is a factor of $\approx 10$ greater than the density inferred in \citet{2013ApJ...770..118M} and is actually quite close to the best-fit profile of \citet{Salem2015RAMMEDIUM}. Some of these best-fit profiles, including the entirety of Table 1 from \citet{2017arXiv171002116L}, are shown for comparison in Figure \ref{SalemRam_hotHalo}, along with the orbital radius, expected velocity, and ram pressure used in our work. The resulting face-on projection of the LMC is also shown in Figure \ref{SalemRam_hotHalo} and is qualitatively very similar to the fiducial stripping results of \citet{Salem2015RAMMEDIUM}, which used all three velocity vector components. 

The inclusion of radiative cooling in our simulations may change the dynamics of the stripping. On one hand, cooling decreases the thermal energy of the initial shock wave and regions heated by the interaction of ram pressure with the galaxy interface, thereby weakening the stripping process. On the other hand, \citet{Tonnesen2009GASMEDIUM} found that cooling actually increased the efficiency of stripping. In their simulations, cooling results in a more complex density structure that allows the surrounding gas to create holes in the disk and strip material from many radii, not just at the leading edge. Cooling also collimates the stripped gas into a narrow wake behind the galaxy instead of flaring it to the sides. This effectively increases the stripping rate. With our model, tests of RPS without cooling result in a stronger initial shock wave and more flared material originating on the leading edge. With cooling, we see more collimated material behind the galaxy, which is consistent with \citet{Tonnesen2009GASMEDIUM} and may have some bearing on filament and Stream formation compared to models without cooling. Models of ram pressure stripping with a magnetized galaxy and IGM suggest that magnetic fields may also alter the shape of stripped gas \citep{2014ApJ...784...75R, 2014ApJ...795..148T, 2017arXiv171101252R}. For now, the inclusion of magnetic fields is left to future work.   

%%%%%%%%%%%%%%%%%%%%%%%%%%%%%%%%%%%%%%%%%%%%%%%%%%%%%

\section{Wind Launching}
\label{windlaunching}

\subsection{Energy and Mass Injection}
Galactic wind simulations in the literature vary in a number of ways, most noticeably in their prescriptions for how supernovae deposit energy and mass into their surroundings and drive outflows. Unfortunately, these processes occur on scales smaller than the finest resolution element used in most simulations, including this present work, which necessitates a ``sub-grid" model. To launch an outflow, we inject both thermal energy and mass into a sphere of affected grid cells. As in \citetalias{2016ApJ...819...29B}, we parameterize mass addition according to a quadratic profile in radius from the center of the injection region:

\begin{align}
q(r) = q_{0}(1-\frac{r^{2}}{R^{2}}) \qquad & \text{for } r < R \\
q(r) = 0 \qquad & \text{for } r > R \\
\end{align}
Here q(r) is the mass per unit time per volume injected into the wind, $q_{0}$ is q(r) evaluated at the center of the injection sphere, and $q_{0}$ is calculated such that $\dot{M} = \int_{0}^{R} q_{0}(1-r^{2}/R^{2}) dV$.
\begin{equation}
q_{0} = \frac{\dot{M}}{\frac{8}{15}\pi R^{3}}
\end{equation}
The mass injection rate, $\dot{M}$, and the hole radius, R, are free parameters. Because individual core collapse supernovae occur only for massive stars with mass greater than $\approx 6-10  M_{\odot}$ and because we find very little difference in the outflow dynamics when we assume 1, 5, and 10 solar masses are deposited per supernova into the surrounding medium (see Section \ref{discussion} for other considerations as well), we choose to inject 10 solar masses per supernova. We then assume that every supernova deposits $10^{51}$ ergs of energy in the gas, and this follows the same parabolic radial profile as the mass injection. The injection is ramped up slowly with a $1-e^{-t/\tau}$ profile with $\tau = 10^{14}$ s $\approx 3$ Myrs. 

Because most of the HI holes of the LMC are along the leading edge at a few kpc from galactic center, we launch each wind 2.5 kpc from galactic center and in the midplane of the disk. Observations reveal a range of HI hole sizes in the LMC \citep{2002AJ....123..255O, 2003ApJS..148..473K} with a mean hole radius $R \approx 100$ pc. We choose this to be our fiducial injection region radius, though we note that the final cavity that is blown out in our fiducial model is a factor of a few larger in radius. Throughout the mass and energy addition, this injection location is tracked as it rotates around the disk at $\approx 40$ km/s to ensure that wind launching follows the site, which can be thought of as a cluster of stars going supernova. 

As noted before, we implement this injection as a source term for thermal energy only, and we neglect any injection of kinetic energy or cosmic ray energy. In reality, some fraction of the injected energy is deposited via all three modes (see e.g. \citealt{Kim2015MOMENTUMMEDIUM}, \citealt{Martizzi2015SupernovaMedium}, \citealt{2016ApJ...826..200S}, \citealt{2017ApJ...834..208R}). The injected thermal energy is subject to radiative cooling, and the energy injection rate must exceed the cooling rate to trigger any energy addition in the affected grid cells. At a radius of 2.5 kpc from galactic center, the gas has a density of $\approx 6 \times 10^{-25} \rm g/cm^{3}$ and temperature of $\approx 2.8 \times 10^{4}$ K. For these gas parameters and the cooling curve used, this implies a minimum supernova rate of $\approx 6$ supernovae per Myr in a 100 pc sphere. For each of the failed outflows labeled ``inadequate heating'' in Table \ref{outflowTable}, the supernova rate was less than this $\approx 6$ supernovae per Myr threshold.

Given uncertainties in how star formation occurs, our inability to resolve individual supernovae, as well as how star formation should be implemented as a sub-grid prescription in simulations, we choose not to use an explicit star formation and feedback scheme. Instead, we leave the supernova rate as a free parameter, but we motivate our choices based on the SFH of the LMC \citep{Harris2009THECLOUD}. Across the entire LMC, the past SFR is of order $0.1 \rm M_{\odot}/yr$. Assuming a conservative SFR of $0.01 \rm M_{\odot}/yr$, which is more appropriate for individual regions of the LMC such as the Southeast Arm, and assuming a Salpeter IMF \citep{1955ApJ...121..161S} for which one supernova occurs for roughly every 166 $M_{\odot}$ of stars formed, this corresponds to a supernova rate of order $10^{-4}$ per year. Over a timescale of a few tens of Myrs, this gives $\approx$ 1000 supernovae. We assume that this outburst occurs over a duration of $\tau = 30$ Myrs, which is an appropriate timescale for stellar cluster evolution. This will be our fiducial model, which we refer to as $\rm SNe1000-lowRes$, $\rm SNe1000-medRes$, or $\rm SNe1000-highRes$ depending on the simulation resolution explained in Section \ref{galaxymodel}. A comparison of this wind at three different resolutions is given in Section \ref{resolution}. 

\begin{table*}[]
\centering
\label{outflowTable}
\begin{tabular}{| c | c | c | c | c | c |}
\hline
Model Name & Injection Duration & SNe Injected & dE/dt (SNe/Myr) & No Ram Pressure & With Ram Pressure \\
\hline
$\rm SNe1000-medRes$  & 30 Myrs & 1000 & 33 & Extended fountain & Expelled \\
$\rm SNe750-medRes$  & 30 Myrs & 750 & 24.75 & Fountain & N/A \\
$\rm SNe500-medRes$  & 30 Myrs & 500 & 16.5 & Small fountain & N/A \\
\hline
$\rm SNe1000-lowRes$  & 30 Myrs & 1000 & 33 & Extended fountain & Expelled \\
$\rm SNe750-lowRes$  & 30 Myrs & 750 & 24.75 &  Fountain & Partially Expelled \\
$\rm SNe500-lowRes$  & 30 Myrs & 500 & 16.5 & Small fountain & Not expelled \\
\hline
N/A (lowRes) & 30 Myrs & 100 & 3.3 & Inadequate heating & N/A \\
N/A (lowRes)  & 30 Myrs & 2000 & 66 & Very large outflow & N/A \\
N/A (lowRes) & 30 Myrs & 10000 (215 pc hole) & 333 & Very large outflow & N/A \\
\hline
\end{tabular}
\caption{Table of injection parameters. Each wind was launched from a 100 pc radius injection, except for the last simulation, for which the total number of supernovae injected was ten times greater and spread out over a volume ten times larger. The last two columns state what type of fountain or wind resulted in isolation and, when applicable, whether that gas was expelled by ram pressure when ram pressure was included. The mass of gas expelled from the disk depends on when the wind was launched, so quantitative values are not included in this table. Instead, the last column gives a general description of whether that outflow launched at a simulation time of $t = 600$ Myrs was expelled by ram pressure.} The top three simulations were run with AMR at a maximum resolution of 19.5 pc (denoted as ``medRes''). These three outflows are shown in Figure \ref{windParam_plot}. With the exception of the ``SNe1000 - medRes'' outflow simulation with ram pressure (see Section \ref{RPS_resTest}, all simulations with ram pressure included (the middle three simulations denoted as ``lowRes'') have a maximum resolution of 78 pc and were run both with and without ram pressure.
\end{table*}

\subsection{Wind Parameter Study}
\begin{figure*}[]
\label{windParam_plot}
\centering
\includegraphics[width=.99\textwidth]{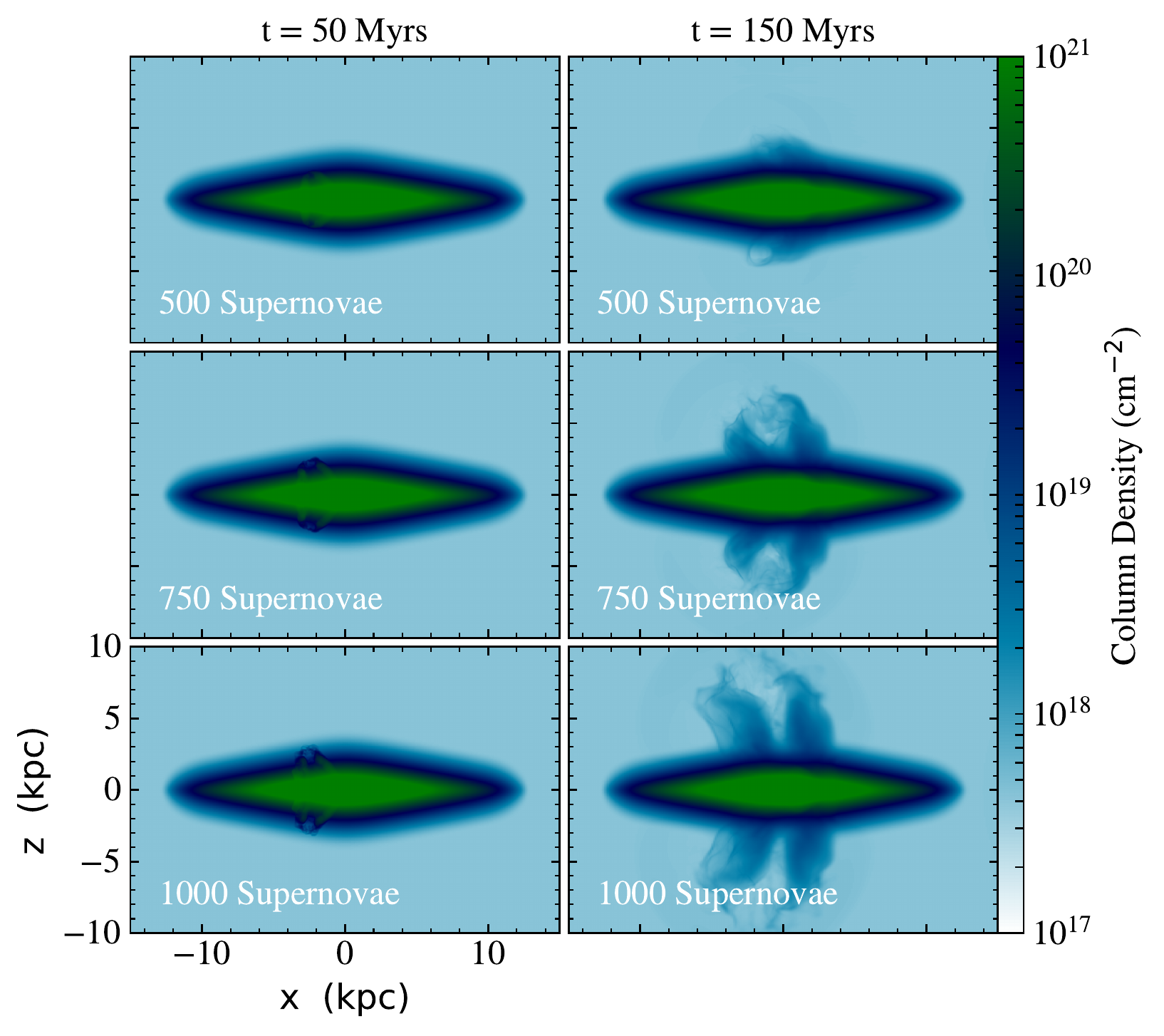}
\caption{The results of three different outflows at various times run at medium resolution. Each wind is launched 2.5 kpc left of center with sustained energy and mass injection for 30 Myrs. Over that time period, 500 supernovae (top), 750 supernovae (middle), and 1000 supernovae (bottom) are injected. These models are denoted as $\rm SNe500-medRes$, $\rm SNe750-medRes$, and $\rm SNe1000-medRes$ in Table \ref{outflowTable}.  At t = 150 Myrs, each outflow has roughly reached it's maximum extent above the disk, but the morphology, mass ejection, and gas velocities are quite different.}
\end{figure*}

Mass and energy are injected at a rate of $\frac{\textrm{\# of supernovae}}{\tau}$. To test how robust our results are, we vary this rate across our suite of simulations by varying the total number of supernovae injected for a given injection duration, $\tau$. In addition to the 1000 supernovae of the fiducial model, we test 500 supernovae and 750 supernovae at a maximum resolution of 19.5 pc. The resulting outflows ($\rm SNe500-medRes$, $\rm SNe750-medRes$, and $\rm SNe1000-medRes$), are shown in Figure \ref{windParam_plot}. Defining the mass of the outflow as the mass difference above 2 kpc between the initial state and the $t = 100$ Myrs state, the outflow masses are $4.17 \times 10^{5}$, $8.64 \times 10^{5}$, and $1.28 \times 10^{6} M_{\odot}$ for 500, 750, and 1000 supernovae, respectively. This does not take into account the mass swept up into a shell in the disk, which may exceed these masses of the ejected component.

The full suite of simulation runs is outlined in Table \ref{outflowTable}, which shows the number of supernovae injected, the time period over which energy was injected, the supernova rate, and a rough description of the result for that outflow model with and without ram pressure. Note that all simulations with ram pressure were run at ``low'' resolution (defined in Section \ref{galaxymodel}), and all of these outflows tested were fountain flows in absence of ram pressure. The last three models display the importance of increasing or decreasing the supernova rate. Below the threshold of $\approx 6$ supernova per Myr, the model with 3.3 supernova per Myr never triggers any energy injection due to radiative cooling. The last two models with 66 and 333 supernova per Myr result in large outflows with significant mass flux off the top and bottom of our simulation box. The last simulation specifically injects the same number of supernova per volume as $\rm SNe1000-lowRes$ but increases the injection radius from 100 pc to 215 pc, resulting in 10 times more energy injected overall.

\begin{figure*}[]
\label{faceon}
\centering
\includegraphics[width = 0.99\textwidth]{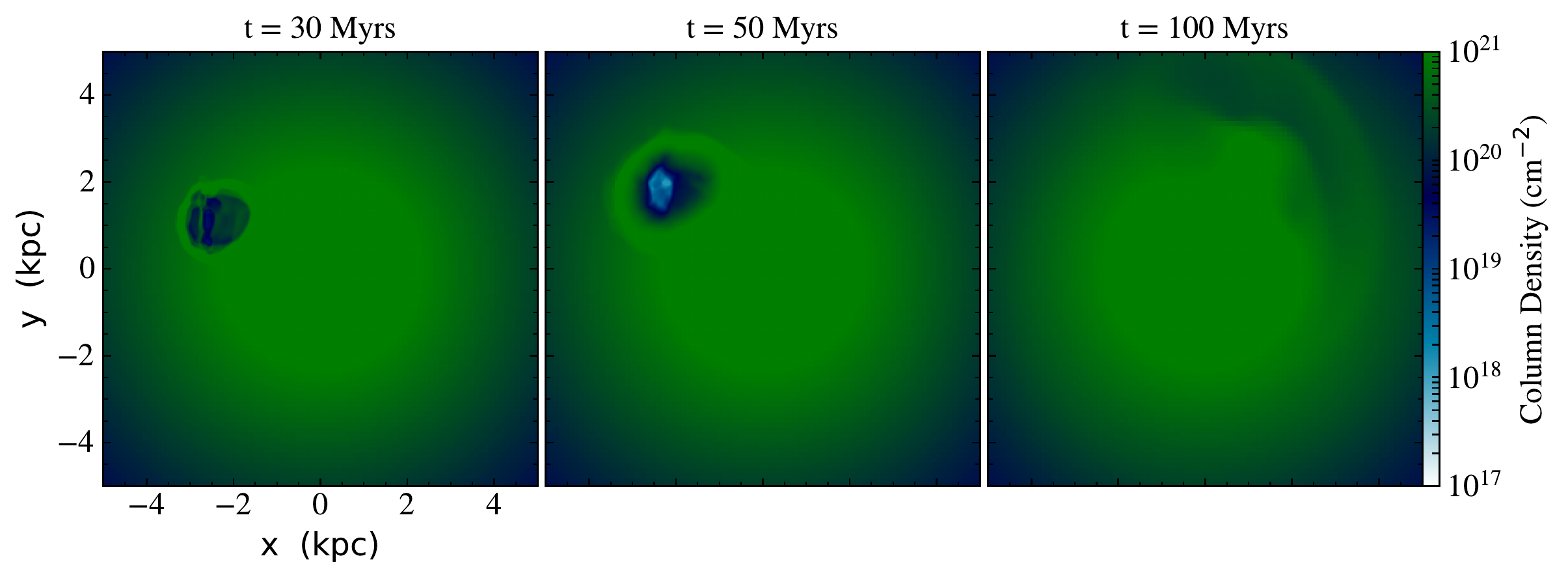}
\includegraphics[width = 0.99 \textwidth]{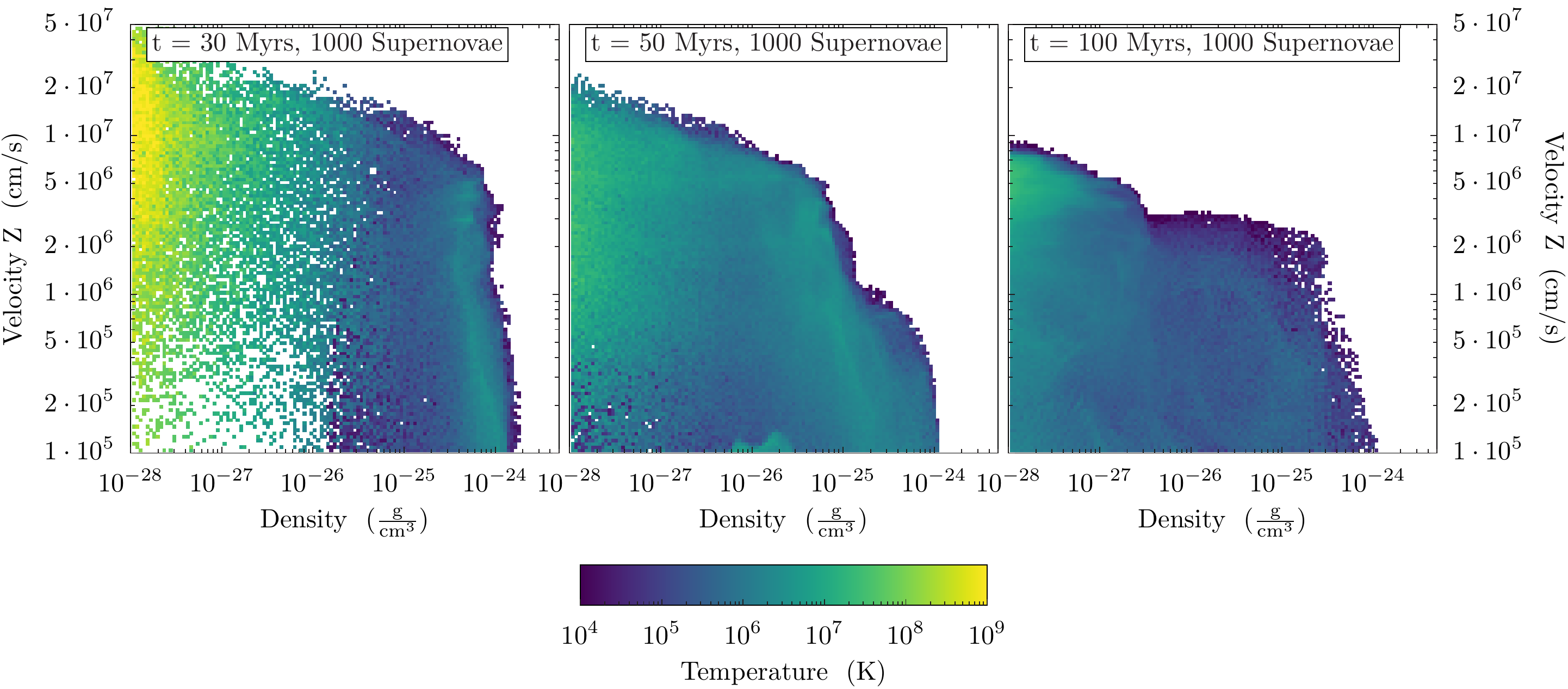}
\caption{Top: Zoomed-in, face-on view of the inner 5 kpc of the galaxy at three different times for the $\rm SNe1000-medRes$ wind: 30 Myrs (left), 50 Myrs (middle), and 100 Myrs (right). At the end of the energy injection period, 30 Myrs, there is a clear hole formed by the pressure difference between the injection region and its surroundings. By 100 Myrs, the hole has collapsed. Bottom: Phase plots for the $\rm SNe1000-medRes$ outflow. The first panel shows low density, hot material as well as high density, colder material being lifted above the disk with velocities of order 100 km/s, which compares well with observations by \cite{2016ApJ...817...91B}. Soon after the energy injection stops, the wind cools to $10^{4} - 10^{5}$ K by a combination of efficient radiative and adiabatic losses, and the gas velocity decreases.}
\end{figure*}

For $\rm SNe1000-medRes$, a hole clearly forms within the first 10 Myrs. The pressure gradient resulting from the increased thermal energy in the hole blows out a ring of material in the midplane, as shown in Figure \ref{faceon}. Since there is less dense material above the disk, the density cavity becomes elongated as material preferentially flows vertically instead of in the midplane. Cold, dense gas forms a shell around hot, low density gas and begins to exhibit the biconical shape characteristic of outflows from disk galaxies (shown in Figure \ref{windParam_plot}). The shape, especially when shown as an edge-on projection, is not truly biconical, however, because the injection site is rotating around the galaxy, so some material rotates towards the trailing side of the galaxy. After a few hundred million years, though, most of the ejected gas falls back onto the disk. Therefore, this outflow, in absence of ram pressure, is really a galactic ``fountain''. 

For 500 supernovae and 750 supernovae (shown edge-on in Figure \ref{windParam_plot}), the results are similar, but the fountains are less extended and fall back to the galaxy on shorter timescales. Lower supernova rates were also tested, but the energy injection rate for many of them didn't exceed the radiative cooling rate; therefore, no energy and mass injection occurred, and no outflow formed. This illuminates radiative cooling as the main challenge facing thermally driven winds. 

To give a sense of the temperature and out-of-plane velocity of this outflow, Figure \ref{faceon} shows phase plots of temperature, z-velocity, and density, as well as a zoomed-in projection of the hot cavity, at three different times. At 30 Myrs, the hole is filled with low density, hot gas on the order of $10^{8} - 10^{9}$ K, and some of the higher density gas swept up by the wind has temperature of order $10^{6}$ K, which may radiate in x-rays. Possible reasons for the high temperature in the cavity are discussed in Section \ref{discussion}. After 30 Myrs, the energy injection turns off. For a short period, the pressure gradient continues to expand the hole, consequently lowering the density even more, as shown in Figure \ref{faceon}. Eventually, the hole collapses, which may lead to further episodic star formation in the disk \citep{1999MNRAS.309..941D, 2004MNRAS.352..363M}, and the wind quickly cools to $10^{4} - 10^{5}$ K. Some low density gas has upward velocity greater than 100 km/s during the wind driving stage, which compares well with observations by \cite{2016ApJ...817...91B}. By 100 Myrs, most of the wind has slowed down considerably.

\subsection{Discussion and Limitations}
\label{discussion}
The small parameter study of thermally driven winds gives a sense of the energy injection required to expel gas above the disk at velocities $\approx 100$ km/s; however, our simplified model of solely thermal energy injection shows a few limitations. For instance, the phase plots in Figure \ref{faceon} show low density gas being heated to about $10^{9}$ K, but these high temperatures are not physically realized. Thermal conduction, which is not included in our model, may diffuse this large temperature build-up, but the diffusion timescale can be quite short and affect the computational tractability of these simulations. An enticing fix is to impose a temperature ceiling, specifically when energy is input into low density regions. However, for our thermally dominated cavity, a temperature ceiling will cap the amount of additional thermal energy input. Imposing temperature ceilings of $10^{8}$ and $10^{9}$ K on our $\rm SNe1000-lowRes$ wind resulted in very different outcomes. 

Limiting the sound speed by adding more mass may be a better option (e.g. \citealt{2014MNRAS.437.3312S}). We tested mass injections of 1, 5, and 10 solar masses per supernova and found that the resulting wind structure and velocity were almost identical between the runs. The only difference was a factor of a few increase in temperature when only 1 solar mass per supernova is used. This makes sense in that $10^{51} $ ergs per 1 $M_{\odot}$ corresponds to gas temperatures of $\approx 2 \times 10^{9}$ K. Given that supernovae eject at least a few solar masses into the ISM, injecting 10 solar masses not only lowers the temperature, which makes the computation more tractable, but also is physically motivated. 

These large temperatures may also result from our inability to properly resolve individual supernovae, which evolve through various phases dependent on the surrounding interstellar medium. At resolutions of order 20 pc and with a mean gas density at 2.5 kpc of $6 \times 10^{-25} \rm g/cm^{3}$, we are not resolving the so-called cooling radius of each supernova remnant, which suggests that most energy injected into the surrounding medium should be in the form of momentum feedback instead of thermal feedback (e.g. \cite{Kim2015MOMENTUMMEDIUM}). If our model included some portion of kinetic energy injection, winds may be launched at lower temperatures, and less of the injected energy budget would be subject to radiative cooling. If another fraction of energy is injected as cosmic rays, this will also help drive a wind. 

Additionally, strong winds from Wolf-Rayet stars may further decrease the susceptibility to radiative cooling by clearing the surrounding medium of dense gas, and overlapping supernovae modeled at higher resolution may boost the effective energy injection. Future work will address these factors and give a more detailed parameter study of wind launching in the LMC, but we believe that the wind prescription used here probably represents the least optimistic scenario for outflow launching \citep{2012MNRAS.426..140D, 2017arXiv170903515S}. Given that local outflows still occur for the reasonable injection rates used here, and given that a more detailed and resolved treatment of wind launching can likely result in similar mass ejections using fewer supernovae, it seems plausible that local outflows from clustered supernovae may pervade the LMC. 

\subsection{Effect of Resolution on Wind Launching}
\label{resolution}

\begin{figure*}[]
\label{resTest}
\centering
\includegraphics[width = 0.99\textwidth]{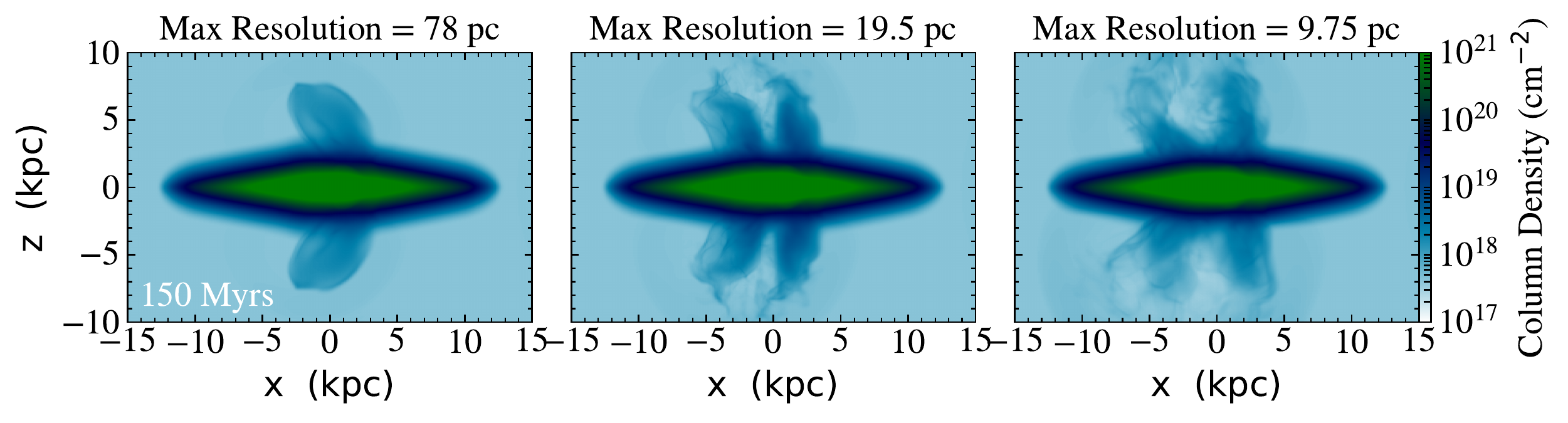}
\includegraphics[width = 0.99\textwidth]{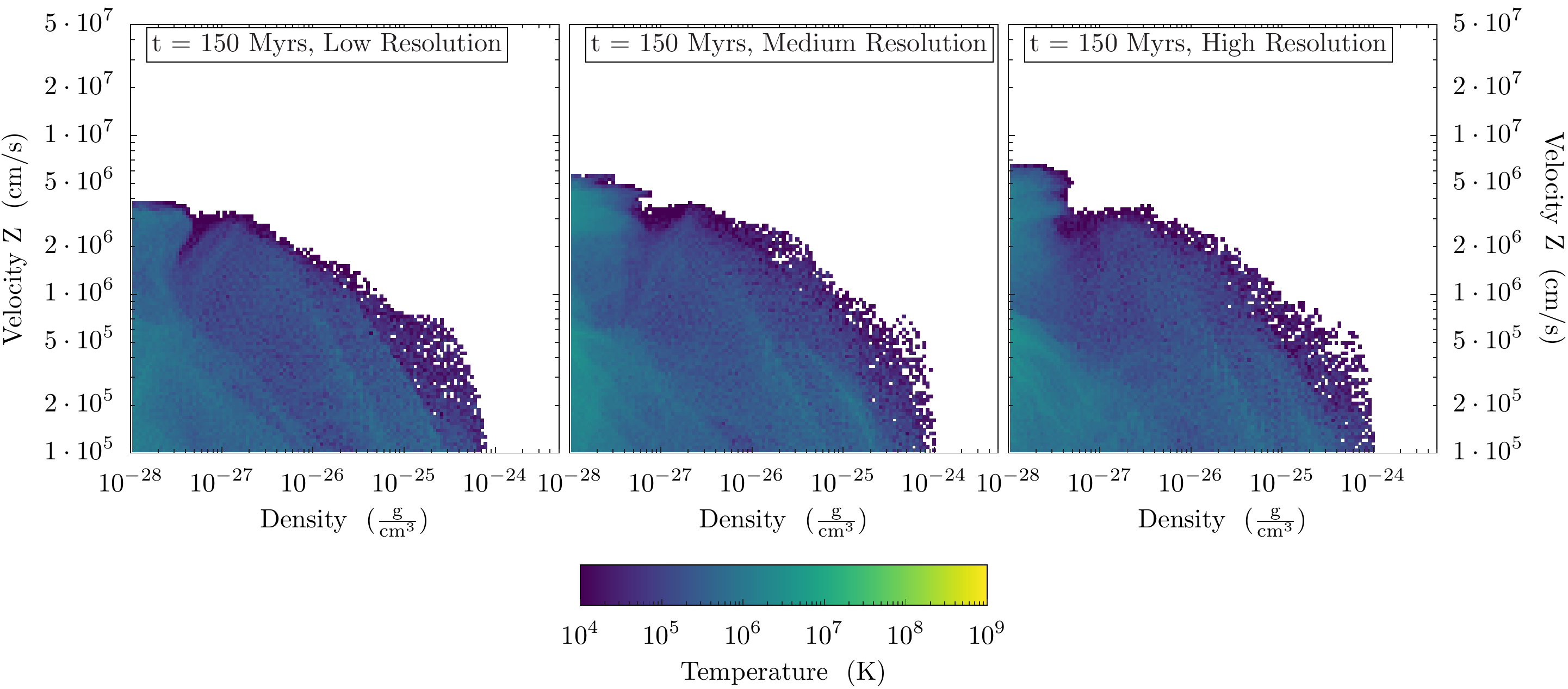}
\caption{Resolution study showing face-on projections (top) and phase plots (bottom) at 150 Myrs for maximum resolutions in the z-direction of 78 pc, 19.5 pc, and 9.75 pc. These resolutions are referred to as ``low'', ``medium'', and ``high'', respectively. The winds are launched in isolation without ram pressure. At $t = 150$ Myrs, the phase plots are very similar, and the wind morphology for the medium and high resolution simulations are similar. The low resolution run results in a smaller fountain flow both in vertical and horizontal extent.}
\end{figure*}

\begin{figure*}[]
\label{resTest_massFlux}
\centering
\includegraphics[width = 0.49\textwidth]{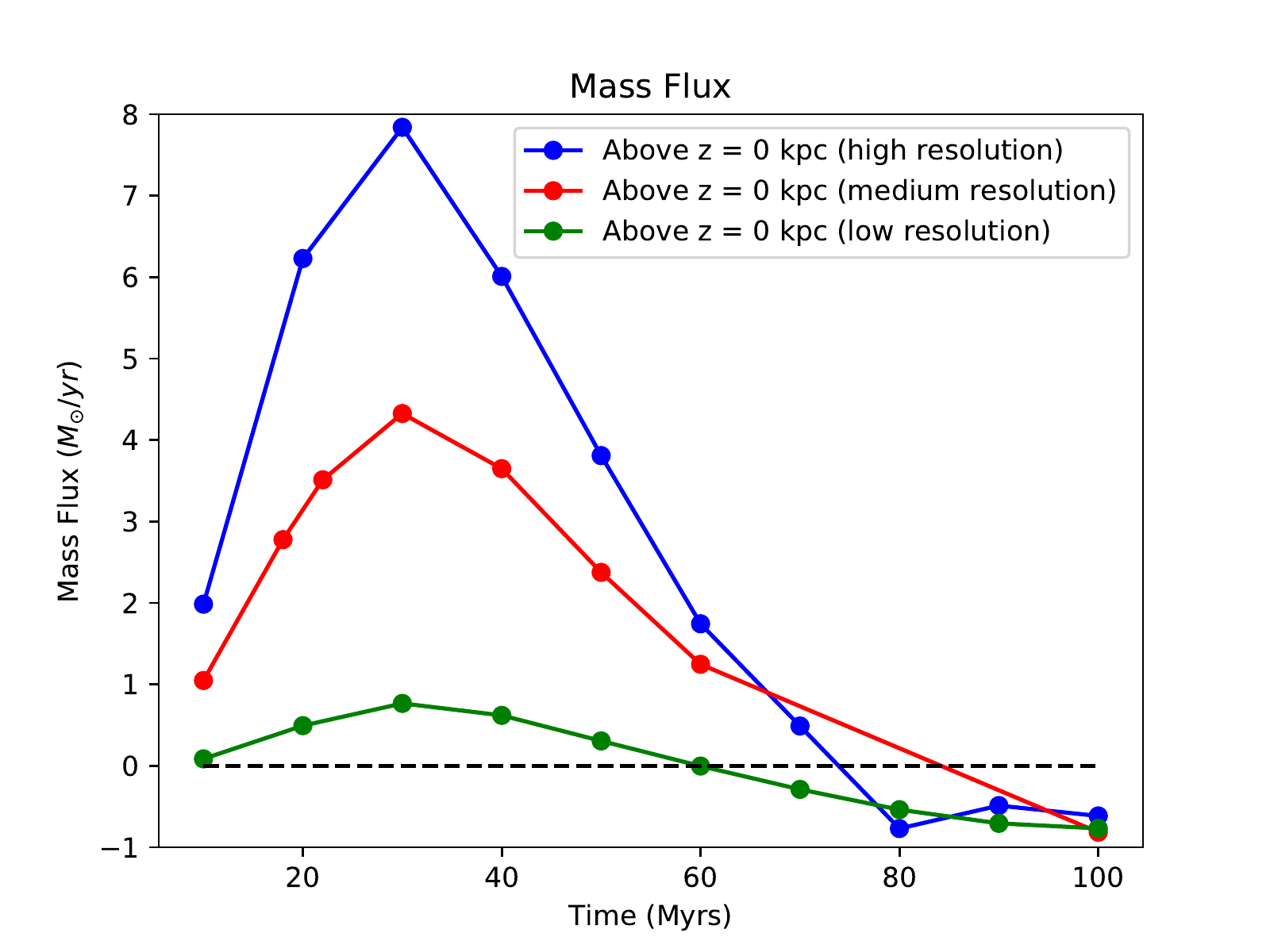}
\includegraphics[width = 0.49\textwidth]{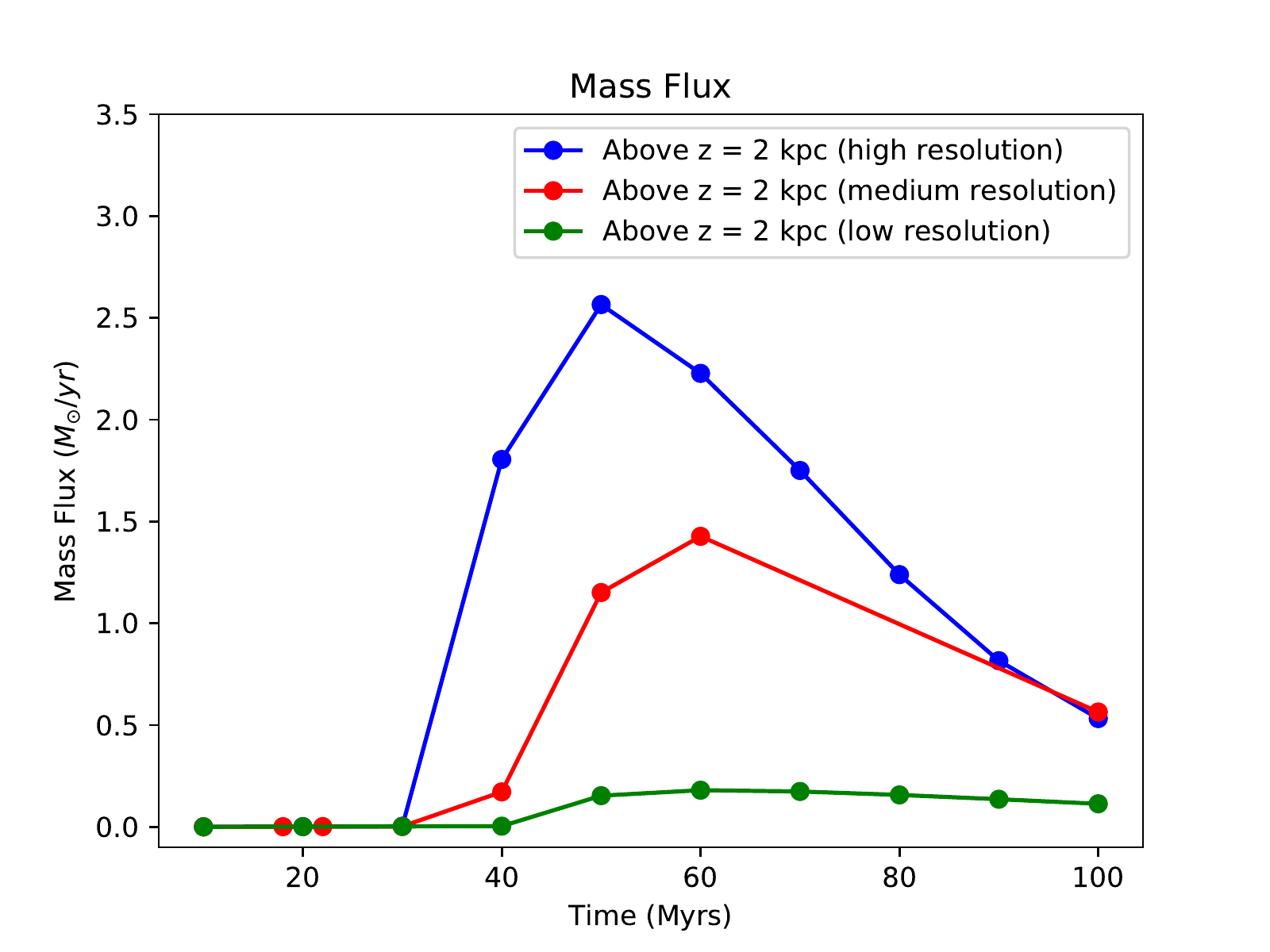}
\caption{Resolution study showing vertical mass flux as a function of time for low, medium, and high resolution SNe1000 simulations. Left: Vertical mass flux throughout the top half of the simulation box ($z > 0$ kpc). Right: Vertical mass flux above $z = 2$ kpc, which isolates the ejected material. The mass fluxes peak at different values depending on resolution, signaling that higher resolution leads to a larger, faster outburst, but the overall behavior is similar. By 100 Myrs, the total mass flux is fairly converged and is negative, signaling that each outflow is turning into a fountain.}
\end{figure*}

Additionally, decreasing numerical diffusion by increasing resolution can change the wind properties. The following resolution study shows the fiducial wind model as it is included in isolation at a highest resolution of 9.75 pc ($\rm SNe1000-highRes$), a medium resolution of 19.5 pc ($\rm SNe1000-medRes$), and at a lower resolution of 78 pc ($\rm SNe1000-lowRes$) as it is for simulations including ram pressure. AMR is utilized in the medium and high resolution runs with 2 levels and 3 levels of refinement, respectively, compared to the low resolution run at base resolution. The grid refines based on temperature, and refinement is triggered only for cells with temperatures above $10^{6}$ K. In the high resolution case, the simulation box contains up to $\approx 250,000$ blocks, with each block containing 8 cells in each direction. 

Higher resolution leads to increased pressure build-up at the wind injection site, which leads to a lower density cavity and more mass ejected above the disk. This is shown in Figures \ref{resTest} and \ref{resTest_massFlux}. The medium resolution and high resolution runs are not entirely converged, though the fountain morphology, spatial extent, and phase plots are roughly the same after 150 Myrs, and the vertical mass flux is comparable after 100 Myrs. The highest resolution run launches gas vertically at a higher velocity than the medium resolution run, leading to vertical mass fluxes almost a factor of two higher than the medium resolution run and slightly higher temperatures. Defining the ejected mass as the mass above $z = 2$ kpc, Figure \ref{resTest_massFlux} shows a peak vertical mass flux in the fountain at $50 - 60$ Myrs in both high resolution and medium resolution runs. Including the disk (the ``above $z = 0$ kpc'' runs), the total vertical mass flux peaks at 30 Myrs when energy injection stops, at which point the pressure differential decreases and the thermal driving subsides. By 100 Myrs, the ejected gas approaches zero mass flux as the wind turns into a fountain, and the total vertical mass flux is slightly negative. 

For the purposes of this pilot study, we consider our medium resolution runs to be at a sufficiently high resolution to draw conclusions on the general shape and characteristics of each outflow in our small sample. The low resolution runs are sufficient for studying the interplay between the outflow gas and ram pressure (see Section \ref{RPS_resTest} for a resolution study); however, they are not resolved enough to relate the outflow morphology and mass outflow rates to the energy injection parameters. As to which resolution leads to the most accurate results, we note that the artificially high numerical diffusion of the low resolution runs in some ways mimics thermal conductivity, which lowers the maximum temperature to more reasonable values. Also, \cite{2016ApJ...817...91B} report a mass outflow rate of $\approx 0.4 M_{\odot} / yr$ from the LMC, which is closer to that of our low resolution outflows; however, that reported rate may be just a fraction of the \emph{total} outflow rate, which we report for our simulations, because only a subset of ionization states is probed by the observations. Given the computational expense of the high resolution runs, it would be imprudent to use our simplified, parabolic thermal energy injection profile in such simulations. At resolutions of order 10 pc, we can begin to resolve smaller scale processes that may alter the wind launching. This will be the subject of future work, as well as a more detailed comparison to observations.

\section{Wind Launching with Ram Pressure Stripping}
\label{windandram_section}

\begin{figure*}[]
\label{ram_noram}
\centering
\includegraphics[width=0.99\textwidth]{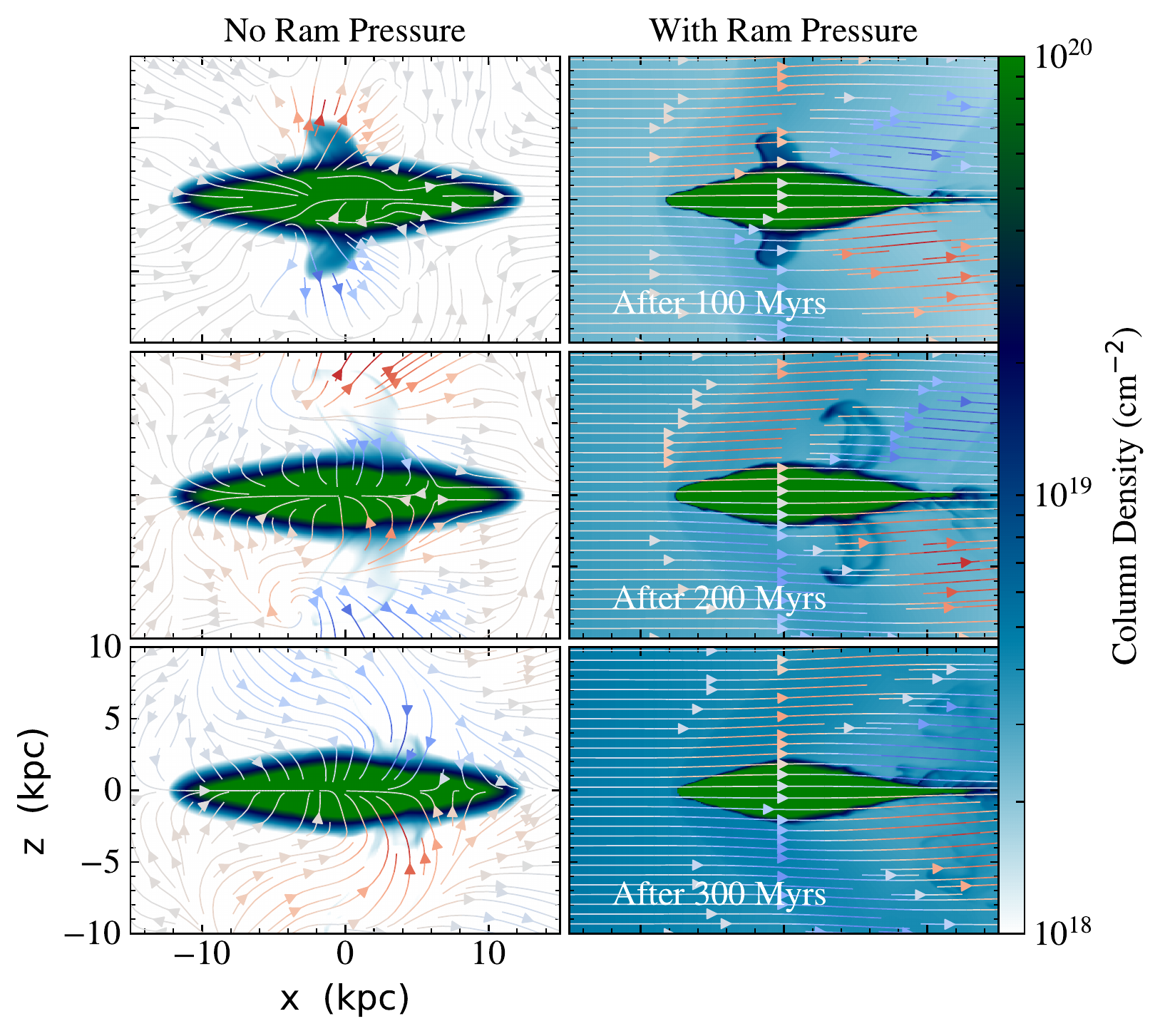}
\caption{Comparison between $\rm SNe1000-lowRes$ outflows launched in isolation and outflows affected by ram pressure after 100 Myrs (top), 200 Myrs (middle), and 300 Myrs (bottom). The wind launch occurs at 600 Myrs in the ram pressure simulation. Velocity streamlines are overplotted with color representing velocity in the $\hat{z}$ direction -- red is positive, and blue is negative. In isolation, the $\hat{z}$ velocity changes sign when fountain gas falls back to the galaxy. When ram pressure is included, the $\hat{x}$ velocity dominates. This results in the fountain flow being converted into expelled material. By 300 Myrs, the wind material is about to flow off the right edge of the grid.}
\end{figure*}

\begin{figure*}[]
\label{ram_noram_750}
\centering
\includegraphics[width=0.99\textwidth]{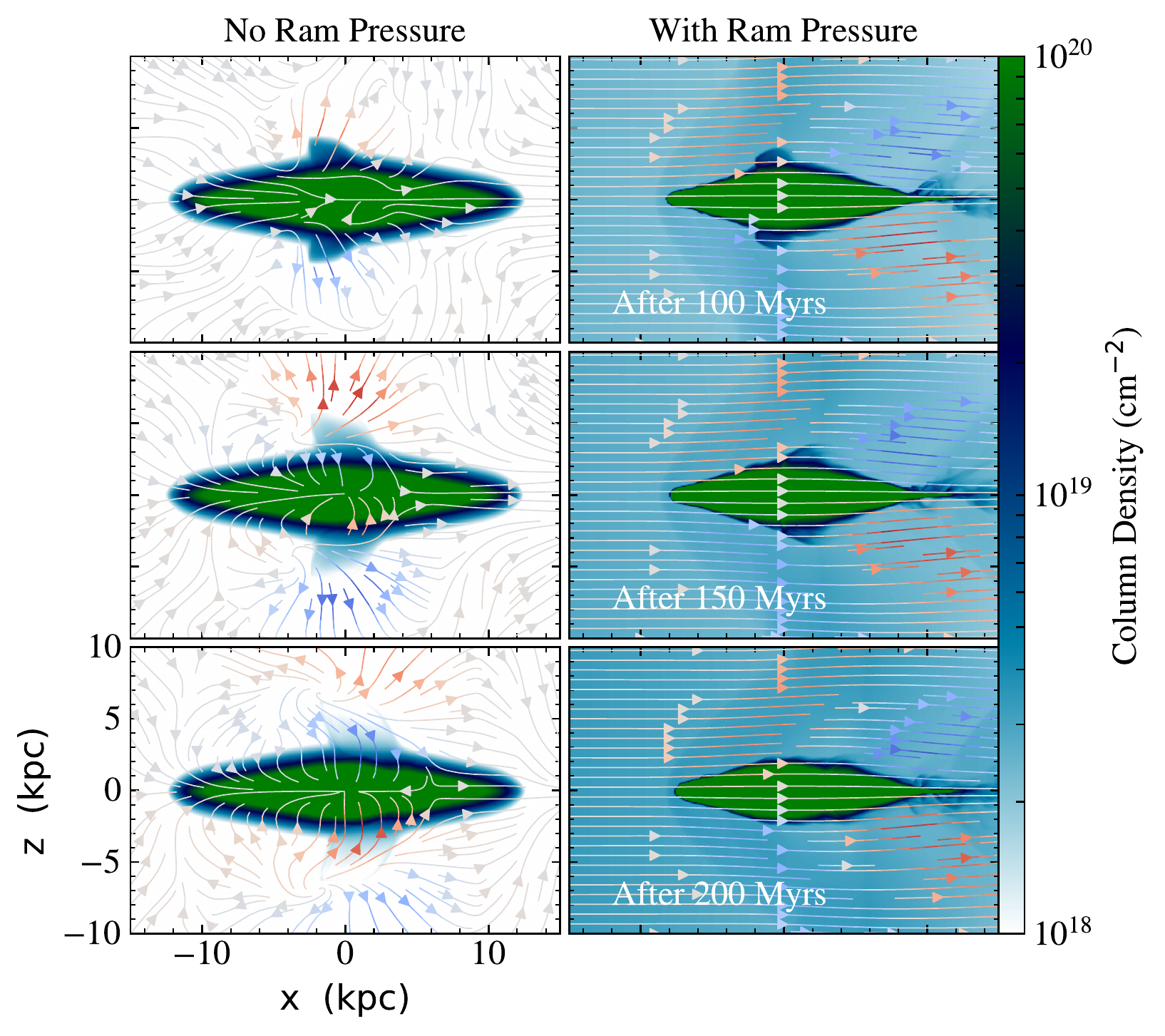}
\caption{Comparison between $\rm SNe750-lowRes$ outflows launched in isolation and outflows affected by ram pressure after 100 Myrs (top), 150 Myrs (middle), and 200 Myrs (bottom). The wind launch occurs at 600 Myrs in the ram pressure simulation. Velocity streamlines are overplotted with color representing velocity in the $\hat{z}$ direction -- red is positive, and blue is negative. Unlike the expelled $\rm SNe1000-lowRes$ fountain, this outflow doesn't achieve the same spatial extent above the disk, and ram pressure doesn't have enough time to expel more than just a small fraction of gas from the galaxy. }
\end{figure*}

In this section, we present results of our simulations with both wind launching and RPS. Here, the wind parameters are consistently those of $\rm SNe1000-medRes$; however, the maximum resolution of 78 pc is lower than that used in the wind parameter study. We name this outflow model ``$\rm SNe1000-lowRes$.'' Because we are primarily interested in whether ram pressure can convert a fountain flow into expelled gas, it is only important to get gas above the disk regardless of how the wind launching is done.  

Figure \ref{ram_noram} shows the outcome of the $\rm SNe1000-lowRes$ outflow with and without ram pressure stripping included. When launched in isolation, the bulk of the outflow, which lifts $\approx 10^{6} M_{\odot}$ of gas above 2 kpc, begins to fall back after about 200 Myrs. The overplotted velocity vectors show very low density gas swirling and mixing with the background higher above the disk while the high density gas close to the disk splashes back onto the trailing side of the galaxy. By 300 Myrs, almost the entirety of the outflow has fallen back.

With ram pressure included starting at $t = 0$, the wind is launched at $t = 600$ Myrs, and the fountain's evolution is shown at 100, 200, and 300 Myrs after being launched, which corresponds to simulation times of $t = 700, 800,$ and $900$ Myrs. Comparing the top two panels of Figure \ref{ram_noram}, one can see the ram pressure compresses the outflow along the leading edge (with a corresponding pressure and temperature increase not shown). After 200 Myrs, the gas forms a bow-shape as it is pushed to the trailing edge of the galaxy. The maximum height of outflowing material above the disk is shortened compared to the isolated wind test; this is partly because the ambient Milky Way halo density is now higher than in the isolated test (which used $10^{-29} \rm g/cm^{3}$) and partly because the stripped, high pressure material now residing above the LMC limits the superbubble breakout \citep{1999MNRAS.309..161M, 2001MNRAS.324..191S, 2004MNRAS.352..363M}. Even with the outflow extent shortened, since the velocity is dominated by the ram pressure in the x-direction, the gas doesn't have enough time to fall back to the disk before the sufficiently high inertia of the ram pressure sweeps it away. At 300 Myrs, the gas can be faintly seen moving off the edge of the grid. This represents an efficient conversion of fountain material into expelled gas. 

\begin{figure}[]
\label{diskMass}
\centering
\includegraphics[width=0.49\textwidth]{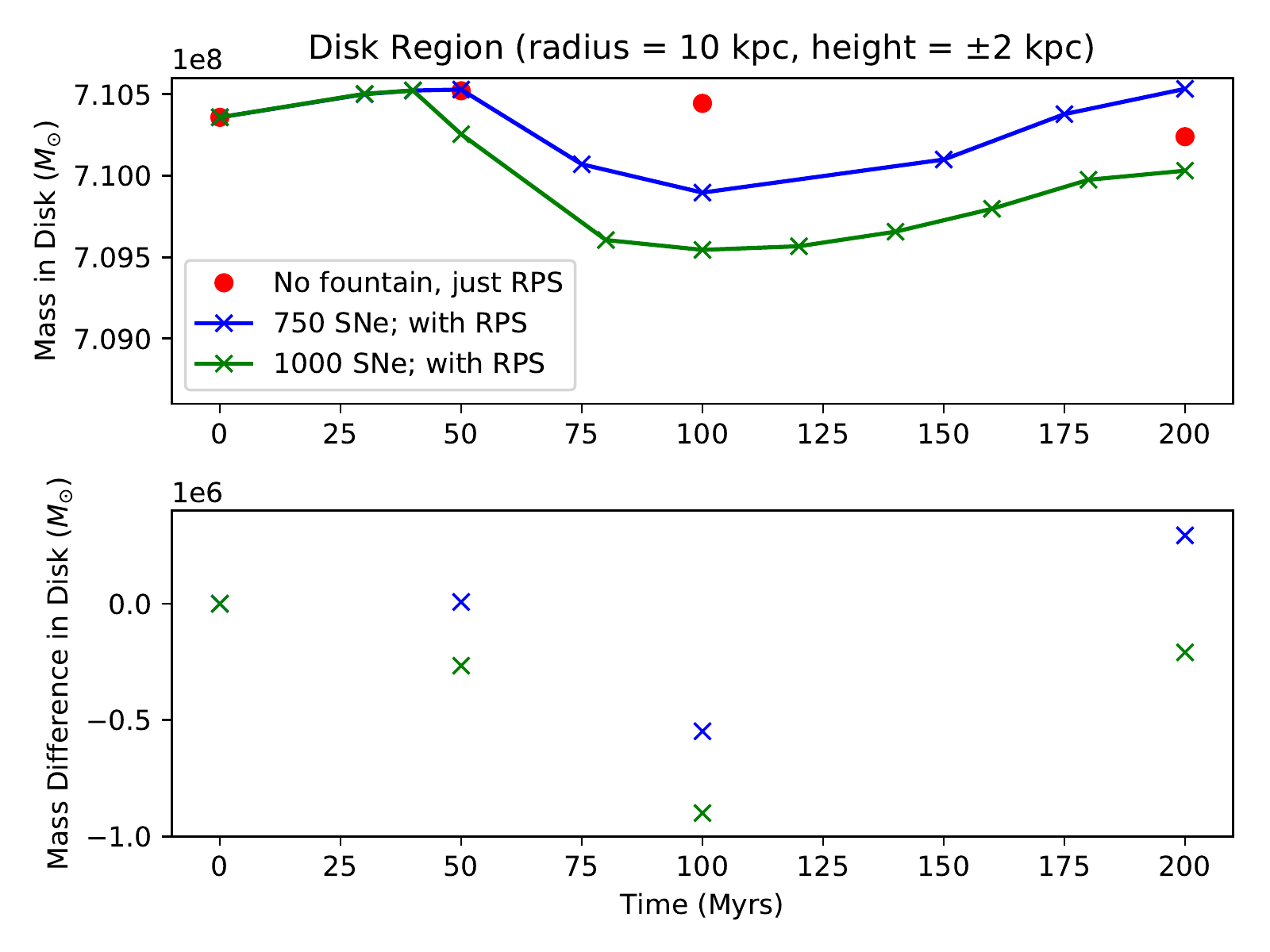}
\caption{Disk mass, defined as the mass within a cylinder of radius 10 kpc and height 2 kpc above and below the galaxy midplane, as a function of time for simulations with just RPS, with 750 supernovae and RPS, and with 1000 supernovae with RPS. The mass difference between the fountain simulations and the RPS only simulation is also shown. This difference due to the fountain is much greater in the SNe1000-lowRes run than in the SNe750-lowRes run. The 750 supernovae case, after 200 Myrs, has a similar disk mass (in fact, slightly higher) to the RPS only case, which is due to the fountain splashing back onto the galaxy. The 1000 supernovae case, though, successfully expels of order a few $\times 10^{5} M_{\odot}$ of disk gas that doesn't fall back.}
\end{figure}

This process doesn't work for all fountains, however, due to a competition between fall-back time and acceleration time by ram pressure. Figure \ref{ram_noram_750} shows the $\rm SNe750-lowRes$ outflow with and without ram pressure included. Here, the spatial extent of the outflow above the disk isn't great enough, and the fountain mostly falls back to the disk before ram pressure can sweep it away. This is shown more quantitatively in Figure \ref{diskMass}. The amount of mass in a disk of radius 10 kpc and height 2 kpc above and below the midplane, centered on galaxy center, is shown as a function of time for simulations with only RPS, with 750 supernovae and RPS, and with 1000 supernovae and RPS. The mass ejected by the outflow, calculated as the difference between the fountain simulations and the RPS only simulation, show that much more gas is lost (and doesn’t fall back) in the SNe1000-lowRes run than in the SNe750-lowRes run, where the amount of disk gas is again comparable to the RPS only gas within 200 Myrs of the outflow launch. The result -- whether the fountain is mainly expelled or falls back to the disk -- seems to depend sensitively on the number of supernovae driving the outflow.

The fate of this gas can have a number of implications for star formation and galaxy dynamics. Specifically, for the LMC, there is recent evidence for a young star population and star formation in the last 50 Myrs on the western (trailing) edge of the LMC. This could be caused by stripped material or galactic fountain material relocating from the eastern to western side of the disk due to a combination of galaxy rotation and ram pressure, thereby providing extra fuel for star formation \citep{Piatti2017StarCloud, 2017arXiv171004053P}. In addition, without ram pressure, angular momentum must be conserved, and the radii at which material can fall back is limited to the inner few kpcs. With ram pressure included, however, material can fall back at greater radii or be completely expelled, acting as an angular momentum transport mechanism.

\subsection{Implications for the LMC Filament}
In regards to the Magellanic system, we do not expect to form the entire Stream from fountain flows. To account for the entire mass budget of the Stream, which is greater than $10^{9} M_{\odot}$ when including both HI and HII, hundreds of small outflows would be necessary. Rather, we are interested in whether one or a few outflows may blow behind the disk and show a filament when projected face-on. As explained further in Section \ref{introduction}, this analysis is motivated by the growing evidence from kinematics and chemistry \citep{2006MNRAS.371..108C, Nidever2008TheArm, 2000AJ....120.1830G, 2013ApJ...772..111R} that show an LMC origin for some of the Stream gas - in contrast to tidal models that typically predict material to be only stripped from the SMC \citep{2012MNRAS.421.2109B, 2014MNRAS.439.1948Y}.

\cite{Nidever2008TheArm} find that the Stream gas is moving away from the LMC at $\approx 49$ km/s, which implies that outflows must have begun ejecting material 1.7 Gyrs ago to account for the full extent of the Stream. Because the ram pressure was very weak that far in the past, and because it is unclear what would have triggered such star formation at that time, it is unlikely that outflows could have contributed significantly to the Stream without significant help from some other process, such as tidal stripping, which is not modeled here \citep{2006MNRAS.369.1021M}. To this end, our wind launching times are motivated instead by the recent SFH of the LMC, as also assumed by \cite{2004A&A...423..895O}. We inject the $\rm SNe1000-lowRes$ wind 200 Myrs, 400 Myrs, and 600 Myrs in the past, which corresponds to times of $t = 800$ Myrs, $t = 600$ Myrs, and $t = 400$ Myrs in our simulations. These injection times are meant to provide a general idea of the timescale needed for a wind to form a filament; however, they are loosely motivated by the SFH of the LMC, which has a peak around 570 Myrs ago \citep{Harris2009THECLOUD}.

By launching winds during various phases of stripping, we also gain some intuition on the robustness of this process, which depends on the ram pressure $\rho v^{2}$ being large enough to accelerate the fountain gas before it falls back.  For this LMC-specific ram pressure profile, each $\rm SNe1000-lowRes$ fountain flow launched at different times is eventually expelled from the galaxy, which suggests that similar ``recently'' launched outflows will likely be swept away by ram pressure.

\begin{figure*}[]
\label{ramandwind_edgeon}
\centering
\includegraphics[width=0.99\textwidth]{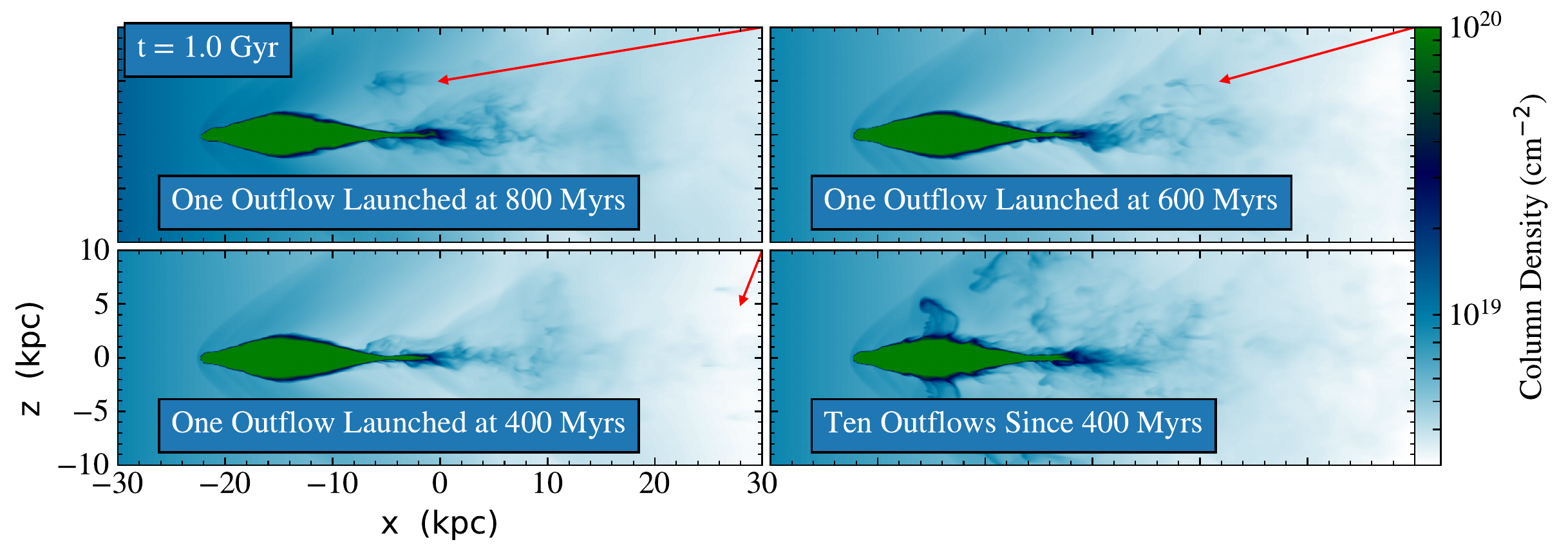}
\includegraphics[width=0.99\textwidth]{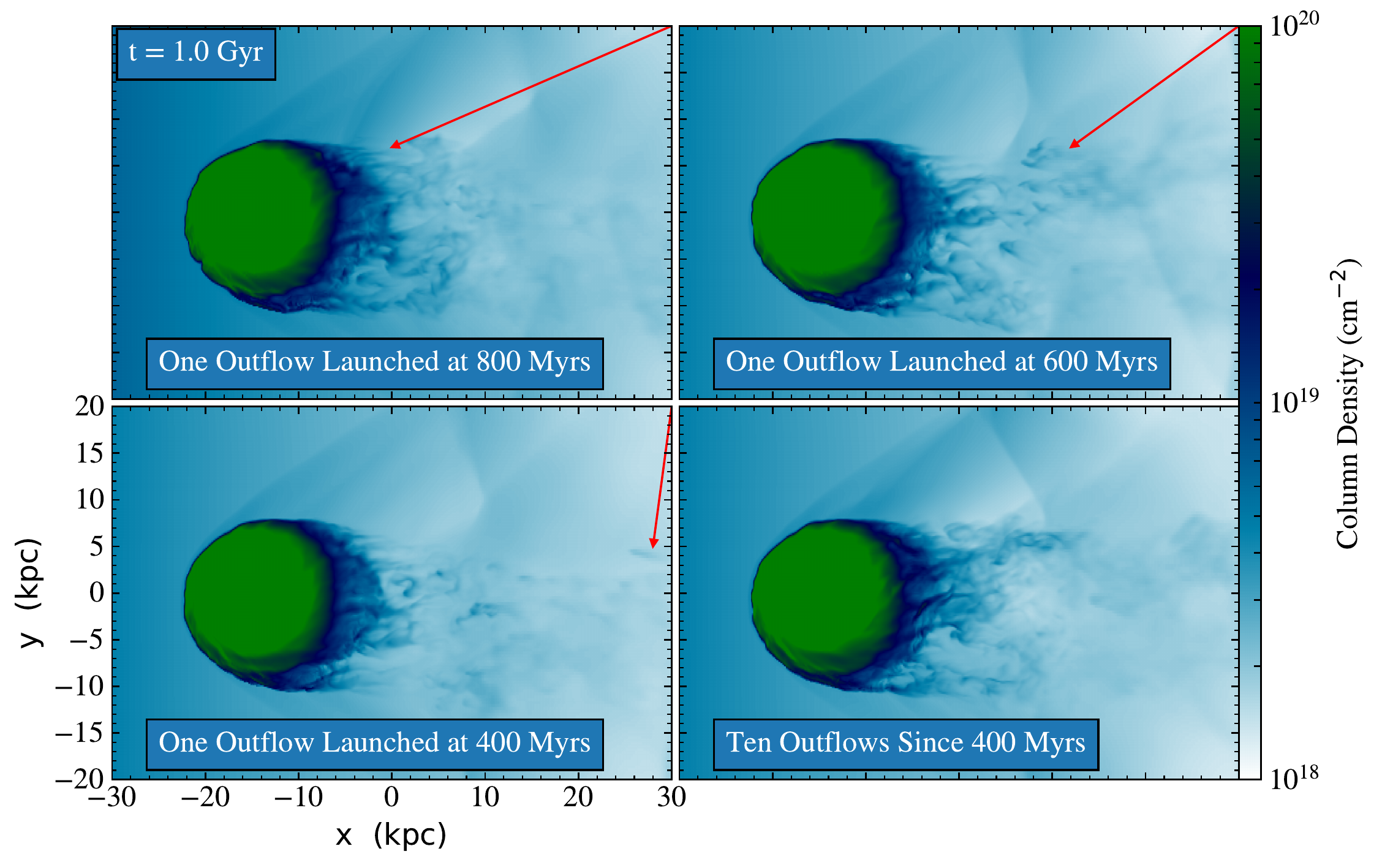}
\caption{Present day, edge-on (top) and face-on (bottom) projections of $\rm SNe1000-lowRes$ winds affected by ram pressure. Individual winds are launched at three different times -- $t = 800$ Myrs (top left), $t = 600$ Myrs (top right), and $t = 400$ Myrs (bottom left). The ``ten outflows'' simulation (bottom right) launches multiple outflows starting at $t = 400$ Myrs. Red arrows point to the general position of the expelled fountain gas. For the multiple outflow case, fountain gas extends from the galaxy center to the edge of the simulation box. When launched at $t = 800$ Myrs, the wind only has 200 Myrs to blow out and be swept up by ram pressure. This isn't enough time to fully strip the galaxy of that gas and form an extended filament; however, a blob of gas does appear behind the trailing edge of the galaxy. When launched at $t = 600$ Myrs, the expelled material disconnects itself from the disk by about 20 kpc, and when launched at $t = 400$ Myrs, the outflow material has been swept more than 30 kpc behind the galaxy and has flowed off the simulation grid. These outflow positions can be more clearly seen in the projections of Figure \ref{projectionCut}, which cut out the inner 5 kpc vertically. The mass outside of this cut region is shown as a function of x-coordinate in Figure \ref{mass_profiles}.}
\end{figure*}

\begin{figure*}[]
\label{projectionCut}
\centering
\includegraphics[width=0.99\textwidth]{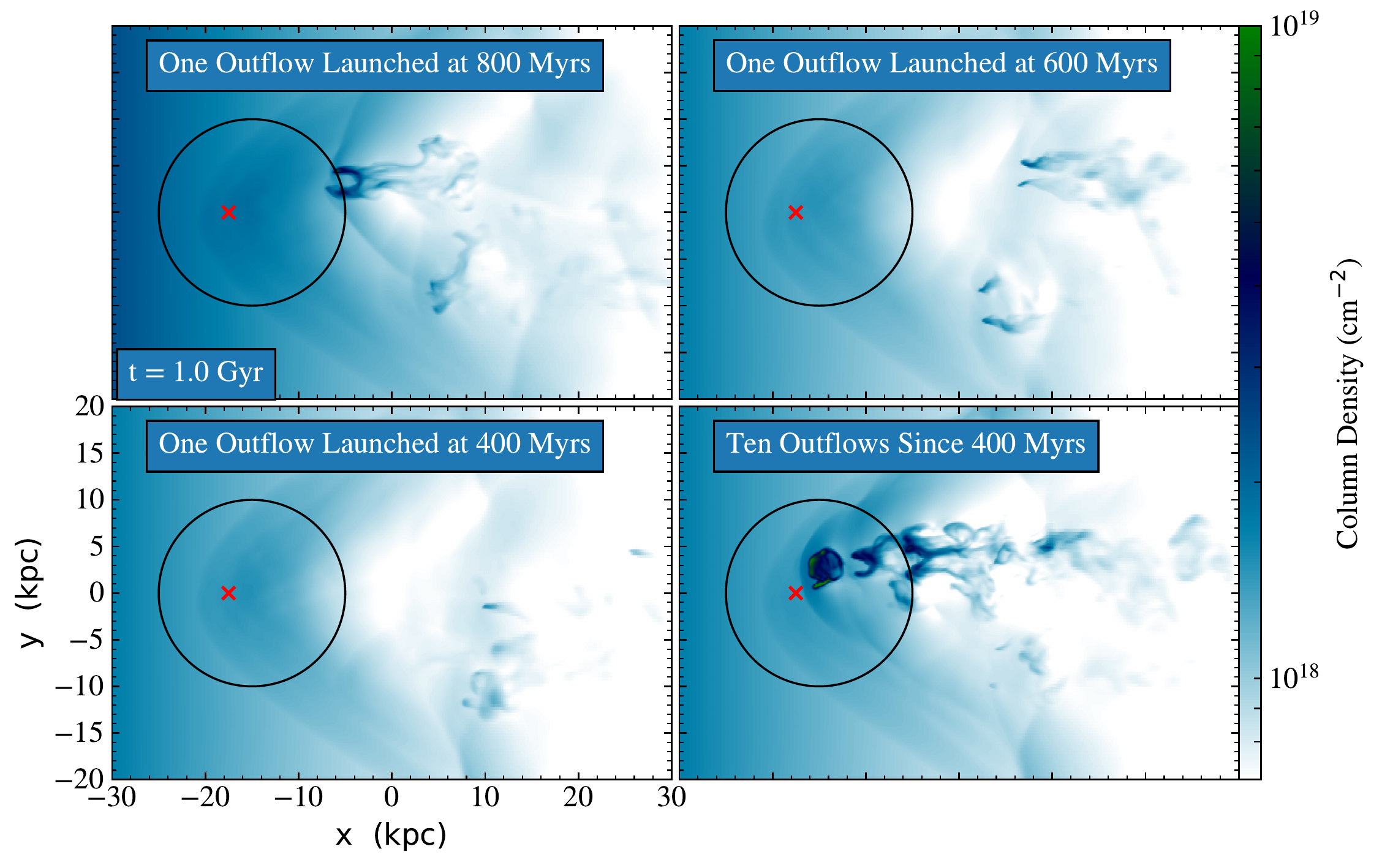}
\caption{Face-on density projections only through the regions $|z| > 2.5$ kpc. This cuts out contributions from the galaxy and RPS tail and isolates the outflow material, which is above the midplane. The red ``x'' marks where the fountains were launched. The black circles of radius 10 kpc show the original extent of the disk before stripping.}
\end{figure*}

Present day ($t = 1.0$ Gyr) edge-on and face-on simulation projections are shown in Figures \ref{ramandwind_edgeon}. In these plots, density is integrated through the entire simulation box in the $\hat{z}$ (face-on) or $\hat{y}$ (edge-on) direction. Edge-on projected density shows that the contribution from outflows is a few kpc above the disk midplane, whereas a clear RPS tail, comprised of non-outflow gas, lies close to the midplane. Red arrows point to the fountain component, which is otherwise difficult to distinguish from the background. To this end, we also plot face-on density projections through only the region $|z| > 2.5$ kpc in our line-of-sight integration. Therefore, we cut out contributions from the galaxy and RPS tail and isolate only the outflow gas. These are shown in Figure \ref{projectionCut}. Red markers indicate where the fountains were originally launched from, and the overlayed circle radius 10 kpc shows the initial extent of the disk before being stripped. This shows how the leading edge has been compressed and also where the outflow gas lies in relation to the disk.

\begin{figure}[]
\label{mass_profiles}
\centering
\includegraphics[width=0.49\textwidth]{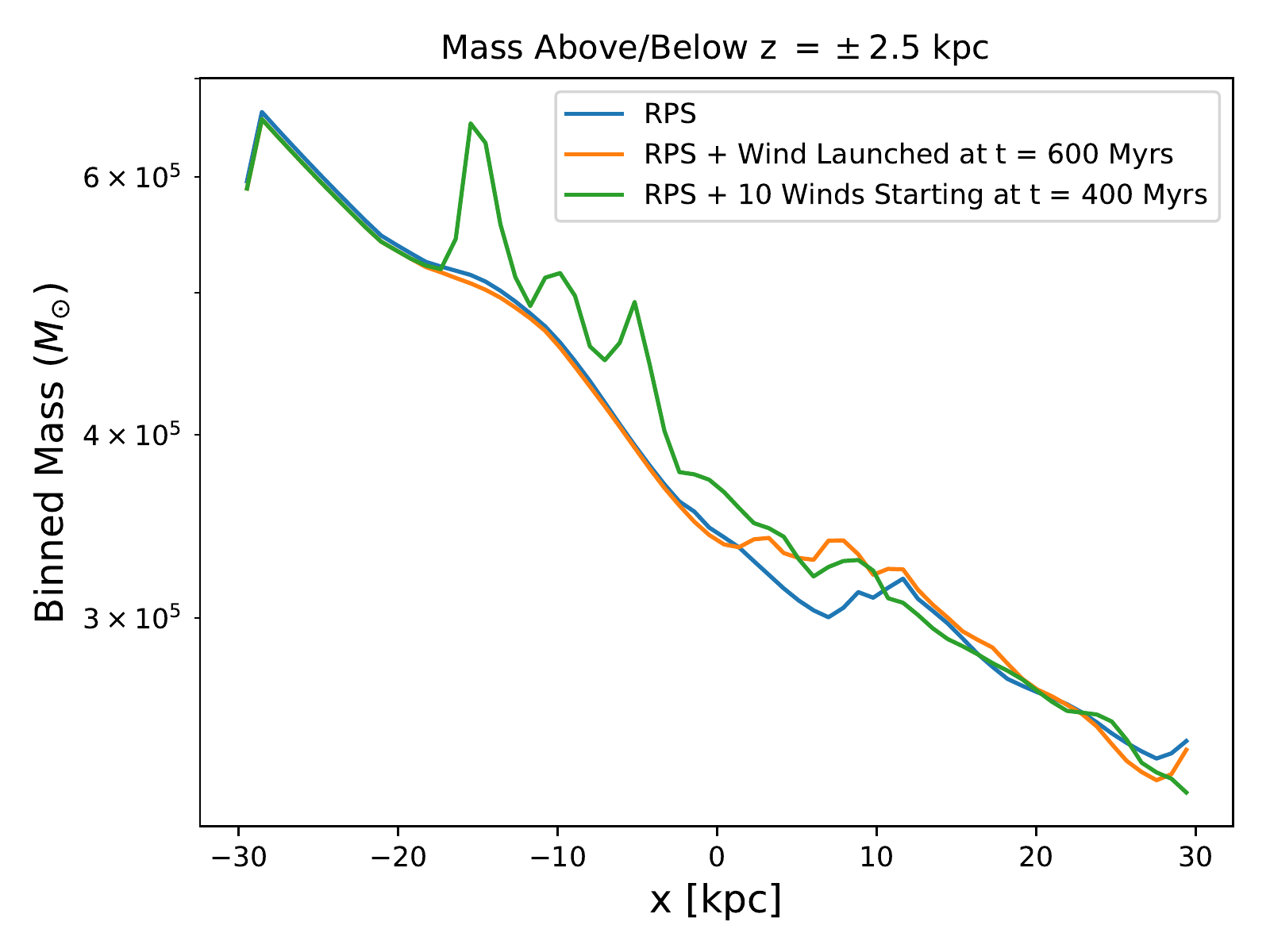}
\caption{Present day (simulation time $t = 1.0$ Gyr) mass in the regions $|z| > 2.5$ kpc as a function of x for simulations with the ``SNe1000-lowRes'' outflow launched at $t = 600$ Myrs, ten outflows starting at $t = 400$ Myrs, and RPS only. The center of the LMC is at $x = -15$ kpc, and the grid is split into 64 equally spaced bins in the x-direction. Using the RPS only simulation as a baseline, the orange bump in mass near $x = 10$ kpc is the single outflow that has been swept downstream 20-30 kpc over the course of 400 Myrs. The same outflow launched at $t = 400$ Myrs is not shown because it has flowed off the grid edge at $x = 30$ kpc. The many green bumps are the various outflows launched at 60 Myr intervals starting at $t = 400$ Myrs. Outflows launched more recently (shown closer to LMC center) are concentrated, dense spikes, while outflows launched earlier have mixed with the background as they are swept to greater x.}
\end{figure}

For the wind launched at $t = 800$ Myrs, the edge-on projection clearly shows the ejected material blowing towards the back side of the galaxy; however, in only 200 Myrs until present day, the ram pressure doesn't have much time to carry the ejecta very far beyond the trailing edge of the disk where the filament is observed. This suggests a timescale limitation for this ejecta-formed filament scenario to be plausible: winds launched within the last $100-200$ Myrs will likely not contribute to the RPS tail. Launching a wind at $t = 600$ Myrs and allowing it to move over the disk for 400 Myrs results in a blob of material that separates itself by about 20 kpc from the disk edge. The same wind launched at $t = 400$ Myrs shows that the expelled blob has moved at least 30 kpc behind the galaxy and has mostly flowed off the simulation grid. 

This one-outflow case serves as a proof of concept that a galactic fountain, launched near the peak in star formation from a localized region of the LMC, can be carried from the galaxy by RPS and contribute, to a fair spatial extent, to the Stream and/or LMC filament. This begs the question, then: can multiple, temporally separated winds lead to a train of connected blobs behind the galaxy, resulting in a filament-like structure?
 
To test the effects of ram pressure on multiple outflows, we successively launch the low-resolution $\rm SNe1000-lowRes$ outflow 10 times starting at $t = 400$ Myrs, with intervals of 60 Myrs between each wind launching. These intervals are long enough for the pre-existing hole to rotate away from the initial injection site before launching a wind there again. Edge-on and face-on results are included in Figure \ref{ramandwind_edgeon}. In this case, the outflows form a clear tail of material when viewed face-on. At column densities close to $5 \times 10^{18} \rm cm^{-2}$, this filament extends roughly 20 kpc behind the galaxy, with lower density wind material extending to the simulation boundary. These column densities, as for the individual outflow simulations, are only comparable to the RPS tail, which is below the observed column densities of the Stream. 

To get a more quantitative sense of the mass ejected from the disk and its spatial position at present-day, we again compare the mass within a cylinder of radius 10 kpc and height 2 kpc above and below the midplane for simulations with one outflow, multiple outflows, and RPS alone. The difference in final mass (at 1 Gyr) of the disk due to the outflows is consistently 3-4 $\times 10^{5} M_{\odot}$ for single outflows compared to RPS only and $3.6 \times 10^{6} M_{\odot}$ for the case where we launch ten outflows in succession. 

The positions of the additional outflow components can be seen in Figure \ref{mass_profiles}. Using the same cuts shown in Figure \ref{projectionCut}, the mass in 64 equally spaced spatial bins is plotted as a function of x-coordinate for the outflow launched at $t = 600$ Myrs and the ten outflows starting at $t = 400$ Myrs, corresponding to the upper right and lower right panels of Figures \ref{ramandwind_edgeon} and \ref{projectionCut}. The RPS only simulation is used as a baseline. The outflow launched at $t = 600$ Myrs can be seen as a mass increase at $x \approx$ 10 kpc, which is 25 kpc from the LMC center. The outflow launched at $t = 400$ Myrs is not shown, as it has already been swept off the grid at $x = 30$ kpc. The multiple outflow simulation shows this full progression of fountain material: recently launched outflows still near the LMC center appear as concentrated spikes of high-density gas, whereas outflows launched further in the past have been swept behind the galaxy to large x values and mixed in with the background medium, showing a more extended mass distribution.

An additional interesting observation is that the first of ten outflows is accelerated by ram pressure more slowly than just an individual outflow launched at 400 Myrs. Subsequently, the filament that forms doesn't move quite as far from the galaxy as the individual outflow. This is because the continuous expulsion of gas above the disk creates a shielding effect, whereby the outflows on the leading edge take on the bulk of the compression from the ram pressure, while the pressure is lower on the previously ejected material. 

This shielding doesn't matter too much for our current LMC simulations because the Milky Way halo density, combined with the LMC's increasing velocity as it falls into the halo, gives a strong enough ram pressure to sweep away multiple outflows. In other systems, such as starburst galaxies moving through the intracluster medium, or if more frequent and massive fountains are launched from the LMC, shielding may play a more prominent role. To exaggerate this effect, we ran the same suite of simulations with a constant halo density of $10^{-29} \rm g/cm^{3}$, which is roughly the density of the Milky Way at 250 kpc from the Galaxy center. We use the same increasing velocity profile, but the present-day consequence of the lower density (shown in Figure \ref{wrongRam}) is apparent. One outflow launched at $t = 400$ Myrs clearly gets pushed behind the galaxy, but multiple outflows shield each other and mainly stay lofted above the disk even after 600 Myrs (at $t = 1.0$ Gyr). With more time, some of this gas will be swept away, but the gravitational field will also pull a larger fraction of it back to the disk. 

\subsection{Resolution Effects}
\label{RPS_resTest}
\begin{figure}[]
\label{lowResmedRes}
\centering
\includegraphics[width=.49\textwidth]{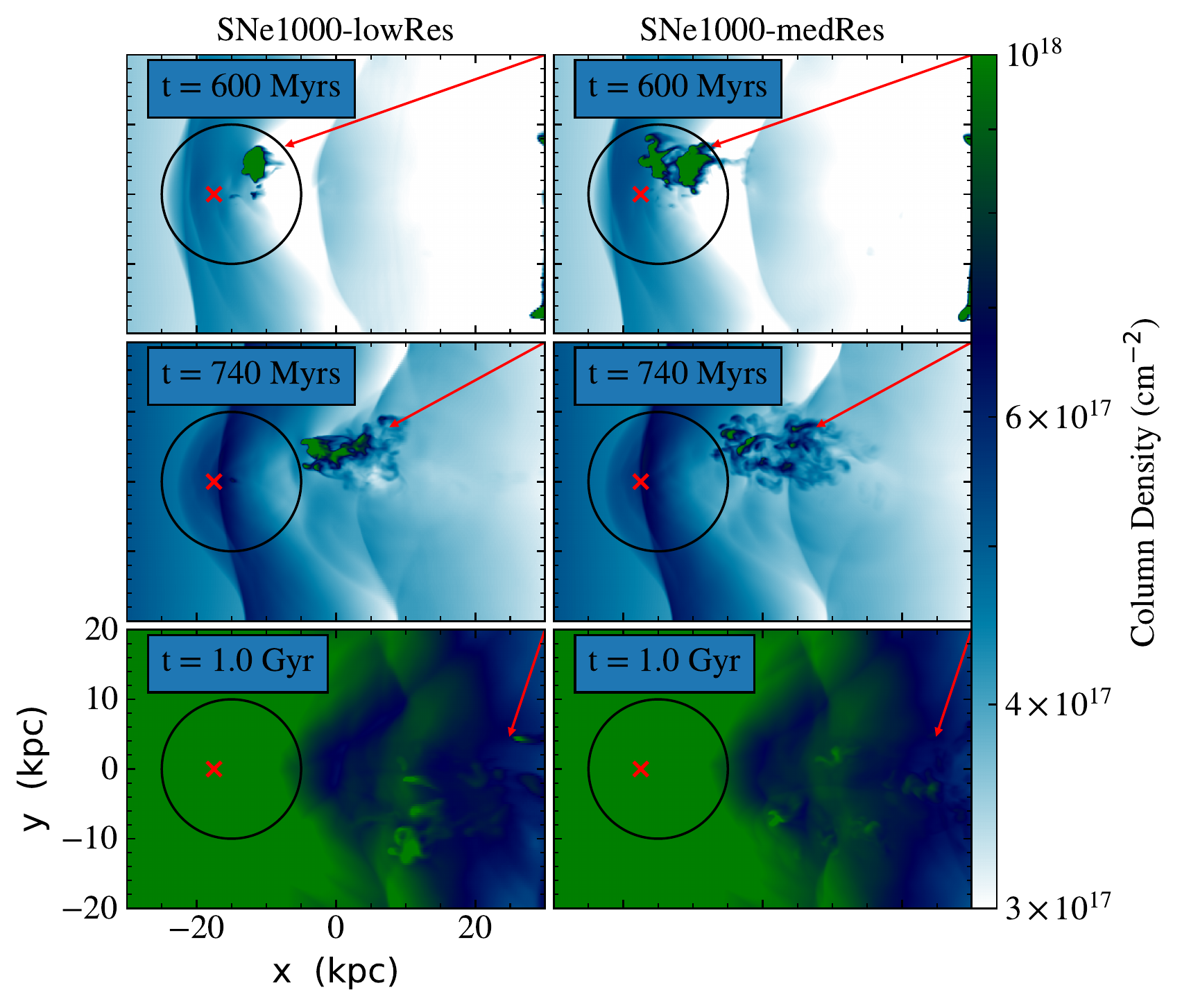}
\includegraphics[width=.49\textwidth]{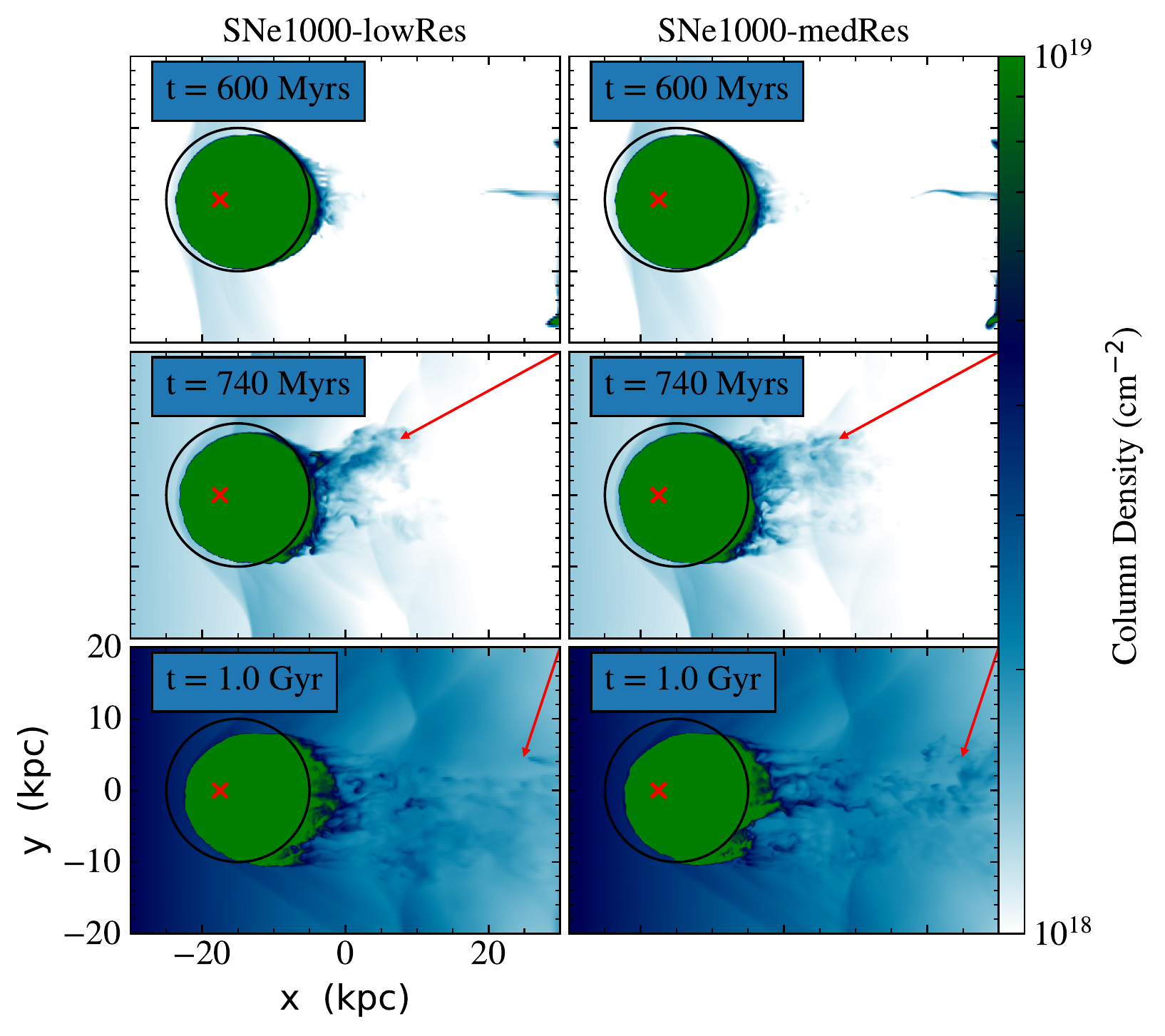}
\caption{Comparing the SNe1000-lowRes (left) and SNe1000-medRes (right) simulations including RPS. Face-on projections cutting out the inner 5 kpc are shown (top) as well as face-on projections through the full volume. The higher resolution run lifts a factor of a few more gas above the disk, and the resulting outflow is more biconical and spread out, resulting in a less concentrated column of gas above the disk. By 740 Myrs, gas is being swept away to a comparable extent, but the higher resolution fountain takes up more volume. By present-day, fountain gas has mostly advected off the simulation box with the red arrows pointing to a barely visible clump of fountain gas exiting the grid.}
\end{figure}

\begin{figure}[]
\label{wrongRam}
\centering
\includegraphics[width=.49\textwidth]{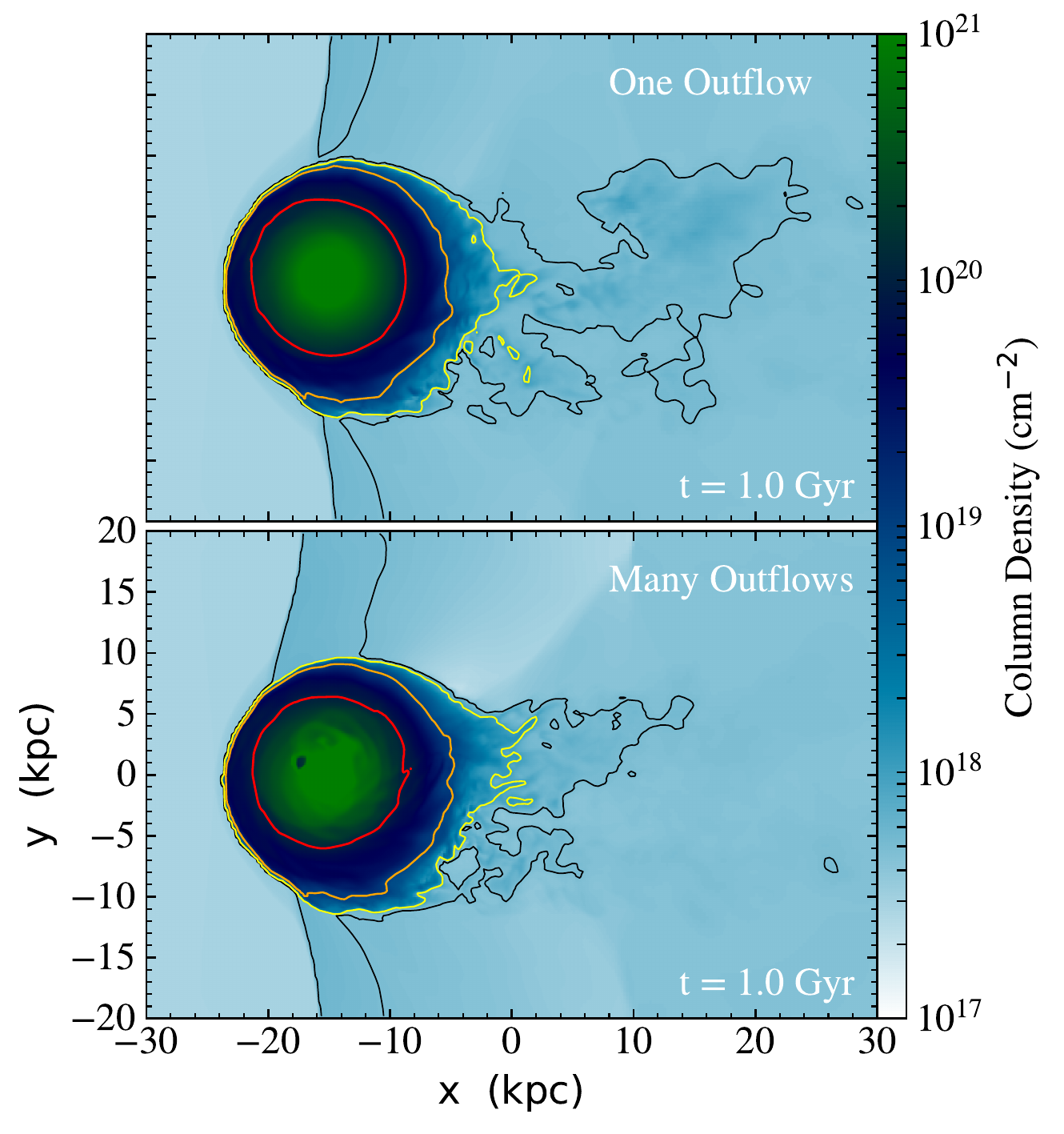}
\caption{Results using the same velocity profile but a \emph{constant density} of $10^{-29} \rm g/cm^{3}$. Face-on views of one outflow (top) and multiple outflows (bottom) at $t = 1.0$ Gyr after wind launching starts at 400 Myrs. Contours show column densities of $10^{20}$ (red), $10^{19}$ (orange), $10^{18}$ (yellow), and $5 \times 10^{17} \rm cm^{-2}$ (black). Whereas a single outflow is efficiently swept up by ram pressure, the multiple outflow scenario shows less gas being pushed behind the galaxy. This shielding effect may somewhat limit filament formation in the LMC and RPS of galaxies in general.}
\end{figure}

Here, we quickly discuss the effects of using AMR to gain higher resolution in the wind launching region. We ran an additional, higher resolution simulation with the same base level of resolution as before, but with two additional levels of AMR utilized in the wind launching region (this corresponds to the ``SNe1000-medRes'' simulation). Figure \ref{lowResmedRes} shows the ``SNe1000-lowRes'' vs ``SNe1000-medRes'' runs with RPS and the outflow launched at t = 400 Myrs. They show face-on projections at three different times after the launch, both through the full simulation box and through only the regions beyond 2.5 kpc above and below the disk. Qualitatively, the column density figures are similar, but the higher resolution outflow is more biconical, making the fountain component more spread out compared to the low resolution run. The fountain component is swept about the same distance behind the disk. 

Quantitatively, we calculate at $t = 600$ Myrs the mass within a disk of radius 10 kpc and height 2 kpc above and below the midplane for the low resolution and medium resolution runs, as well as the simulation with only RPS. Comparing the mass differences in the disk for each case, the low resolution outflow ejects $4.83 \times 10^{5} M_{\odot}$ more than the RPS only, and the medium resolution outflow ejects $1.25 \times 10^{6} M_{\odot}$ more than RPS only. This factor of 2.59 difference means that more mass per fountain is added to the filament, which helps support the outflow origin of the LMC filament; however, it doesn’t change our conclusion that hundreds to thousands of outflows would need to be present to account for the mass of the entire Trailing Stream.

%%%%%%%%%%%%%%%%%%%%%%%%%%%%%%%%%%%%%%%%%%%%%%%%%%%
\section{Conclusions}
\label{Conclusions}
In this work, we have taken a well-studied galaxy with many observational constraints and conducted a small parameter study of wind launching from star-forming regions. The limited applicability of spherically symmetric, steady-state studies motivated fully 3D simulations of local outflows using the FLASH code. This work provides some ``ground-truth'' on thermally driven winds. Consistent with previous works, we find that radiative cooling is an effective inhibitor of many outflows, but physically motivated supernova rates appropriate for clustered supernovae can lead to outflows with velocities in line with recent observations and mass ejections comparable to the HI blobs likely ejected from SGSs \citep{Nidever2008TheArm}. 

Many of our simulated outflows were in fact galactic fountains, where much of the material falls back to the disk within a few hundred million years. \emph{Our simulations serve as a proof-of-concept, though, that ram pressure resulting from the LMC's orbit within the Milky Way halo can convert fountain flows into expelled gas in cases where the fountain's fall-back time is greater than the time it takes for ram pressure to sweep that material away.} This process may be generalized to systems other than just the LMC, and it may be a mechanism to increase stripping and further pollute the IGM \citep{2012ApJ...761...71Z, 2013MNRAS.433.2749G}. In future work we will consider the implications of these results for the mass, energy balance, and composition of the CGM \citep{2017ARA&A..55..389T}. 

It's especially intriguing that simulations of massive satellite galaxies indicate that even modest ram pressure, which alone would not efficiently strip gas, can trigger star formation in the satellite \citep{1999ApJ...516..619F, 2003ApJ...596L..13B, 2014MNRAS.438..444B, 2017MNRAS.466.1382L, 2018arXiv180203019W}, possibly leading to gas loss via both astration and fountain flows converted to expelled gas. The LMC may be a prime example of ram pressure circumventing its stripping inability by initiating this alternative gas-loss channel. Larger scale versions of this process are possible in galaxies falling into cluster centers. Ram pressure from motion through the intracluster medium may not only strip large amounts of gas, resulting in the long trailing tails typical of ``jellyfish'' galaxies \citep{2014ApJ...781L..40E}, but also trigger star formation and gas inflow to the central black holes \citep{2017Natur.548..304P, 2018MNRAS.474.3615M}, therefore feeding AGN jets that can further expel material and bend in the presence of ram pressure \citep{2017PhPl...24d1402J}

\emph{Multiple outflows that lodge a significant amount of gas above the disk, though, may create a shielding effect that minimizes the effect that ram pressure can have on the fountain gas.} This shielding effect may be especially important in starburst galaxies where a large amount of fountain gas may have too much inertia to be pushed by ram pressure on a short enough timescale to be fully expelled from the galaxy. Additionally, a reservoir of high pressure outflows above the disk may mimic the behavior noted by \citet{1999MNRAS.309..161M} and \citet{2004MNRAS.352..363M}, whereby the stripped gas confines superbubbles in the disk and limits further wind launching.

These simulations are focused, however, on the parameters (ram pressure, disk orientation, density, metallicity, etc.) of the LMC in an attempt to study whether these outflows may comprise part of the LMC filament and the Trailing Arm of the Stream. The current best model of the Stream formation has the LMC and SMC interacting in isolation before a recent infall to the Milky Way \citep{2012MNRAS.421.2109B}. This dwarf-dwarf interaction scenario would preferentially strip material from the lower mass galaxy and therefore would not produce long filaments from the LMC. This has been expanded upon in \citep{2018arXiv180201600P}, which shows that even when increasing the tidal stripping from LMC, the dwarf-dwarf tidal model has difficulties producing such long coherent filaments from the larger of the two galaxies. Further, a tidal origin of the Stream would create both Leading and Trailing Tails such that, if the Trailing LMC filament was produced in a tidal model, we would expect to find an LMC filament in the Leading Arm \citep{2018arXiv180201600P}. Recent measurements of the Leading Arm show gas with a range of abundances - but with mostly low abundance gas that has an SMC origin \citep{2018arXiv180106446F}. This result questions the purely tidal origin of the Stream and indicates there might be a second process at work - such as the outflow model presented here. Previous simulations by \cite{2015ApJ...813..110H} of ram pressure and outflows forming the Trailing Stream, though envisioning a different scenario in which the LMC and SMC collide within last 200-300 Myrs and in which significant outflows are launched over a Gyr in the past, similarly underscores the possible importance of supernova feedback in forming part of the Trailing Stream. 

\emph{Our work suggests that outflows and fountains, launched in succession around the peak star formation time and swept up by ram pressure, may contribute to the observed column density of the Stream and LMC filament.} This conclusion is based on our simulations of weak thermally driven winds/fountains with energy injection motivated by the SFH of the LMC. We find that the resulting mass fluxes, both in isolated disk simulations and in simulations including RPS, depend sensitively on the number of supernovae thermally driving the outflow and the resolution of our simulation box. We expect the outflow generation and interaction with ram pressure to then also depend sensitively on the prescription for mass and energy injection, which in this work, we consider to be the least optimistic prescription due to neglect of momentum input and cosmic ray driving. Higher resolution simulations result in higher single outflow mass-loss rates and consequently contribute more mass to the filament. 

Given the uncertainties in the localized SFH of the LMC, as well as uncertainties in the wind launching, it is possible that larger outflows were launched from focused, intensely star forming regions similar to 30 Doradus. Therefore, a significant outflow contribution to the Stream and LMC filament remains a distinct possibility, but likely requires that large outflows were launched early in the LMC's orbit to give ram pressure enough time to accelerate the dense material. The inertia of the surrounding medium relative to the fountain gas determines whether outflow material gets expelled before falling back to the disk. Shielding, then, can limit this expulsion when many winds are launched.

Future simulations with more detailed treatments of wind launching from star-forming regions may lead to more massive fountains than the thermally driven fountains studied here. Reconciling the assumed supernova rates with the past SFH of the LMC will be an important step in determining how many outflows there may be and how much gas they contribute above the disk and into the LMC filament. A growing collection of LMC observations, combined with increasing knowledge of the SFH of the LMC, makes this galaxy an intriguing test-bed for future galactic outflow simulations. 

\section{Acknowledgements}
The authors would like to thank the referee for  thorough, insightful comments that improved this manuscript. The authors also acknowledge helpful discussions with Gandhari Wattal, Yi-Hao Chen, Bob Benjamin, Chris McKee, Greg Bryan, Mordecai Mac-Low, Hsiao-Wen Chen, and Andrew Fox. The software used in this work was in part developed by the DOE NNSA-ASC OASCR Flash Center at the University of Chicago. The data analysis and visualization utilized the publicly available \emph{yt} software (\cite{2011ApJS..192....9T}) and plotly, an interactive graphing library for Python. This work used the Extreme Science and Engineering Discovery Environment (XSEDE), which is supported by National Science Foundation grant number ACI-1548562 \citep{xsede}. C.B. is supported by the National Science Foundation Graduate Research Fellowship Program under Grant No. DGE-1256259. Any opinions, findings, and conclusions or recommendations expressed in this material are those of the author(s) and do not necessarily reflect the views of the National Science Foundation. ED acknowledges the support and the hospitality of the Center for Computational Astrophysics (CCA) at the Flatiron Institute during the preparation of this work. E.G.Z. acknowledges support from the University of Wisconsin-Madison, NSF grant No. AST-1616037, and the hospitality of the Department of Astronomy and Astrophysics at the University of Chicago.

%\section{Appendix: Resolution Study}
%\label{appendix}

%%%%%%%%%%%%%%%%%%%%%%%%%%%%%%%%%%%%%%%%%%%
\centering
\bibliographystyle{apj}
\bibliography{bibliography}
\hypertarget{label1}{}

\end{document}